\documentclass[acmtog,screen,authorversion]{acmart} 
\AtBeginDocument{%
	\providecommand\BibTeX{{%
			\normalfont B\kern-0.5em{\scshape i\kern-0.25em b}\kern-0.8em\TeX}}}
\setcopyright{acmcopyright}
\copyrightyear{2024}
\acmYear{2024}
\acmDOI{XXXXXXX.XXXXXXX}
\acmSubmissionID{114}
\usepackage{stfloats}
\usepackage{float} 
\usepackage{subfigure}  
\usepackage{algorithm}
\usepackage{algorithmic}
\usepackage{bm}
\usepackage{amsmath}
\usepackage{hyperref}
\usepackage{algorithm}
\usepackage{algorithmic}
\usepackage{framed}
\usepackage{ulem}
\newcommand{\note}[1]{#1}

\newcommand{\mbx}{\boldsymbol{x}}
\newcommand{\mbp}{\boldsymbol{p}}

\newcommand{\mbe}{\boldsymbol{e}}
\newcommand{\mbn}{\boldsymbol{n}}

\newcommand{\mbb}{\boldsymbol{b}}

\newcommand{\mbi}{\boldsymbol{i}}

\newcommand{\ct}{\mathcal{T}}
\newcommand{\mbd}{\boldsymbol{d}}
\newcommand{\mbu}{\boldsymbol{u}}

\begin{document}
	
	\acmJournal{TOG}
	\acmYear{2025} \acmVolume{44} \acmNumber{4} \acmArticle{1} \acmMonth{8} \acmPrice{15.00}\acmDOI{10.1145/3618360}

	\title{Bernstein Bounds for Caustics}
\author{Zhimin Fan}
\email{zhiminfan2002@gmail.com}
\affiliation{
	\institution{State Key Lab for Novel Software Technology, Nanjing University}
	\city{Nanjing}
	\country{China}
}
\author{Chen Wang}
\email{chenwang@smail.nju.edu.cn}
\author{Yiming Wang}
\email{02yimingwang@gmail.com}
\affiliation{
	\institution{State Key Lab for Novel Software Technology, Nanjing University}
	\city{Nanjing}
	\country{China}
}
\author{Boxuan Li}
\email{211250189@smail.nju.edu.cn}
\author{Yuxuan Guo}
\email{221240007@smail.nju.edu.cn}
\affiliation{
	\institution{State Key Lab for Novel Software Technology, Nanjing University}
	\city{Nanjing}
	\country{China}
}
\author{Ling-Qi Yan}
\email{lingqi@cs.ucsb.edu}
\affiliation{
	\institution{University of California, Santa Barbara}
	\city{Santa Barbara}
	\country{United States of America}
}
\author{Yanwen Guo}
\email{ywguo@nju.edu.cn}
\affiliation{
	\institution{State Key Lab for Novel Software Technology, Nanjing University}
	\city{Nanjing}
	\country{China}
}
\author{Jie Guo}
\email{guojie@nju.edu.cn}
\authornote{Corresponding author.}
\affiliation{
	\institution{State Key Lab for Novel Software Technology, Nanjing University}
	\city{Nanjing}
	\country{China}
}
	%
	
	\begin{abstract}
		
		Systematically simulating specular light transport requires an exhaustive search for triangle tuples containing admissible paths. Given the extreme inefficiency of enumerating all combinations, we significantly reduce the search domain by stochastically sampling such tuples. The challenge is to design proper sampling probabilities that keep the noise level controllable.
		Our key insight is that by bounding the irradiance contributed by each triangle tuple at a given position, we can sample a subset of triangle tuples with potentially high contributions. Although low-contribution tuples are assigned a negligible probability, the overall variance remains low.
		
		Therefore, we derive position and irradiance bounds for caustics casted by each triangle tuple, introducing a bounding property of rational functions on a Bernstein basis. When formulating position and irradiance expressions into rational functions, we handle non-rational parts through remainder variables to maintain bounding validity. 
		Finally, we carefully design the sampling probabilities by optimizing the upper bound of the variance, expressed only using the position and irradiance bounds.
		
		The bound-driven sampling of triangle tuples is intrinsically unbiased even without defensive sampling. It can be combined with various unbiased and biased root-finding techniques within a local triangle domain. Extensive evaluations show that our method enables the fast and reliable rendering of complex caustics effects. 
		\note{Yet, our method is efficient for no more than two specular vertices, where complexity grows sublinearly to the number of triangles and linearly to that of emitters, and does not consider the Fresnel and visibility terms. We also rely on parameters to control subdivisions.}
		

	\end{abstract}
	
	\begin{teaserfigure}
		\centering
		\begin{minipage}{\linewidth}
			\centering
			\includegraphics[width=1.0\linewidth]{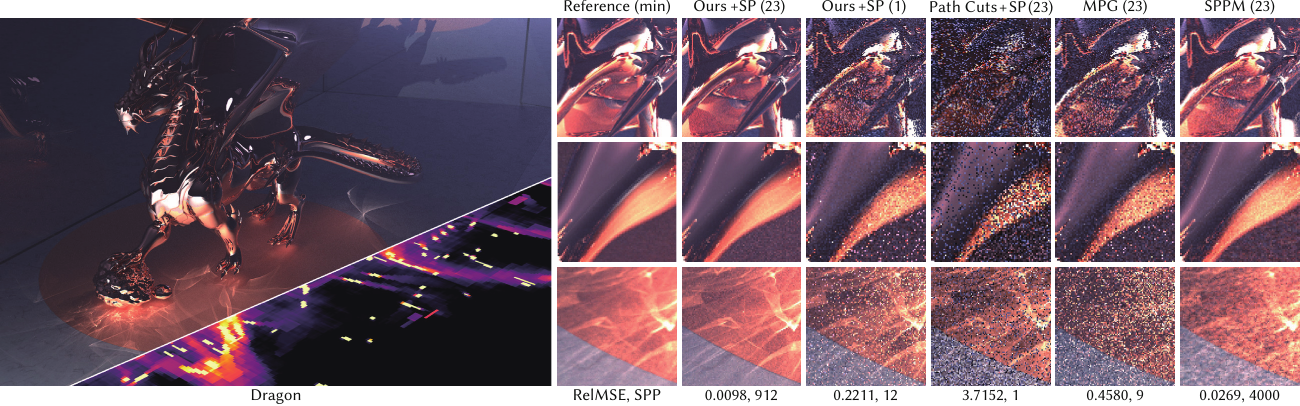}
		\end{minipage}
		\caption{
			Rendering sharp caustics reflected by complex geometry (0.35M triangles), where existing methods perform slowly. Consequently, even if deterministically searching for the complete set of admissible paths, they still produce high variance due to low sample rates. Our method samples triangles leveraging the bounds for caustics, leading to more converged results. We visualize the irradiance bound (in the base 10 logarithmic space) summed over tuples. All methods render single reflections only.
			We compare with Path Cuts \cite{wang2020path}, Specular Polynomials (SP) \cite{fan24}, Manifold Path Guiding (MPG) \cite{fan23mpg}, and Stochastic Progressive Photon Mapping (SPPM) \cite{Hachisuka09SPPM}.
			Two budgets for ours focus on equal time (32 sec for precomputation with finer subdivisions, 23 min in total) and roughly equal quality comparisons (9 sec for precomputation, 1 min in total), respectively.
		}
		\label{fig_teaser}
	\end{teaserfigure}

	

	

	\maketitle

	\section{Introduction}
	
	High-frequency caustics have long been a core challenge in physically-based rendering. To discover the set of admissible paths that satisfy specular constraints, many specialized methods have been proposed \cite{jakob2012manifold, hanika2015manifold}. Generally, all these methods involve a root-finding process to connect a pair of non-specular endpoints (e.g., a point on the light source and a diffuse shading point), but the domain they operate on differs.
	
	While walking on the specular manifold over the whole scene enjoys great generality \cite{jakob2012manifold, hanika2015manifold, zeltner2020specular}, its point-sampling nature makes the convergence hard to guarantee, which could sometimes produce extremely high variance in complex scenes.
	On the other hand, systematic approaches migrate this unbounded convergence by solving for specular paths within each tuple of triangles \cite{fan24, wang2020path, walterSingleScatteringRefractive2009}. Yet, their overall efficiency largely depends on the selection of these triangle tuples. Currently, they rely on interval arithmetics to prune non-contributing regions, which are loose, deterministic, and without energy considerations.
	
	Deterministic enumeration produces pixel-perfect rendering results but incurs significant computational costs. In practice, stochastic sampling is crucial for achieving an efficient and unbiased solution; however, maintaining low variance is essential to avoid noisy results. To facilitate such efficient sampling of specular triangle tuples, it is necessary to gather information on the distribution of irradiance each triangle tuple contributes to the receiver.
	
	Conventional approaches, such as path guiding, fit energy distributions from path or photon samples. \note{Despite their generality and flexibility, they} could encounter many pitfalls. For example, point samples can easily miss certain parts so that the variance could be extremely high. Besides, online training still requires a (usually uniform) initial distribution. To our knowledge, a reliable \note{(i.e., error-bounded)} and self-contained approach for sampling specular paths that enjoys theoretically controllable variance is a clear gap.
	
	Our key insight is that if we can acquire a conservative bound of position and irradiance of caustics cast by each triangle tuple, we can construct a stochastic estimator of the total irradiance at each shading point with controllable variance. The core challenge here is to efficiently obtain a correct bound as tightly as possible.
	
	To this end, we develop a systematic framework to provide the bounds of positions and irradiance along specular paths within a given region. Here, the key ingredient is a rational formulation of vertex positions and Jacobians. We can easily obtain the bounds of their range by expressing these rational functions in the Bernstein polynomial basis \note{\cite{Garloff2012BoundsOT, Narkawicz12}. }
	
	Unlike conventional MC methods on point samples \cite{kajiyaRenderingEquation1986, vorba19, zeltner2020specular}, our approach operates on functions within a finite interval. This may look complex, but arithmetic operations in Bernstein basis can be understood as modeling functions as Bézier surfaces and \note{operating} on their control points \cite{Farouki2012TheBP}, which is explainable, easy to understand, and \note{enjoys} better numerical stability than monomial basis. 
	We finally leverage our triangle sampling to develop a reliable and effective caustics rendering pipeline, which performs \note{up to} an order of magnitude faster than existing \note{unbiased sampling} approaches.
	

	In summary, our main contribution includes:
	\begin{itemize}
		\item A Bernstein bound of position and irradiance for caustics.
		\item A bound-driven sampler with controllable variance.
		\item An efficient pipeline for rendering sharp caustics.
	\end{itemize}
	
	\note{
		Our current method has some limitations. Firstly, the complexity grows rapidly as chain lengths increase. Thus, it is only feasible for one or two bounces. Besides, subdivision is required to achieve reasonably tight bounds when triangles are not small enough, which introduces some parameters. Additionally, the convergence rate of bound tightness is not yet guaranteed.}
	\note{
		In addition, we make simplifying assumptions during derivations. We ignore visibility and the Fresnel term during precomputation. Also, we compute bounds for each triangle tuple, which suffers from complexity growth sublinear to the number of triangles. Lastly, we consider a single, small light source and pure specular surfaces only. Thus, the precomputation cost scales linearly to the number of light sources.
	} 
	
	\section{Related works}

	\note{Advanced sampling techniques, such as Metropolis light transport \cite{veachMetropolisLightTransport1997} and path guiding \cite{mullerPracticalPathGuiding2017, vorba19}, largely accelerate the convergence rate of Monte Carlo rendering. Nevertheless, caustics produced by tiny light sources and specular surfaces still pose a challenge to these local path sampling techniques \cite{Veach98, kajiyaRenderingEquation1986, veachBidirectionalEstimatorsLight1995}.}
	Therefore, \note{productions \cite{Prod2023} have} traditionally utilized photons \cite{Jensen95PM, Hachisuka09SPPM, georgievLightTransportSimulation2012} or regularization \cite{Kaplanyan2013PathSR}. \note{However, despite their generality, these approaches could result in unexpected bias. Thus, researchers have developed specialized methods to unbiasedly estimate those sharp caustics.}

	\paragraph{Deterministic search for specular paths}
	
	\note{Some early approaches perform an exhaustive searching process to find nearly all admissible chains connecting given endpoints.
		Starting from Fermat's principle, interval arithmetic and Newton's method are leveraged to identify reflection paths on parametric surfaces \cite{mitchell1992illumination}. By further introducing the Implicit Function Theorem,} \citet{Chen00} tackles \note{endpoint perturbations of specular paths using high-dimensional Taylor series.}
	In more recent works, for \note{single-refractive specular chains on} triangle meshes, \citet{walterSingleScatteringRefractive2009} introduce a \note{spindle test based on interval arithmetic to prune non-contributing region. Then, they employ (interval)} Newton's method to find admissible paths. \note{While this can be extended to render} multi-bounce glints \cite{wang2020path}, \note{for caustics}, online pruning would experience performance degradation \cite{fan24}, and the interval arithmetic bound could be extremely loose. 
	This challenge motivates us to precompute the bounds of caustics in a tighter manner, narrowing the search domain. 
	Furthermore, as the number of solutions increases, enumerating all solutions is already time-consuming, necessitating the use of stochastic sampling.

	\paragraph{Stochastic sampling on specular manifolds}
	
	Random walks with a Newton solver on specular manifolds enable more robust handling of specular paths for Metropolis sampling \cite{jakob2012manifold, Kaplanyan2014}, which is then applied to a regular Monte Carlo context \cite{hanika2015manifold}. \note{By further introducing random initialization strategies together with reciprocal estimators, unbiasedness is guaranteed \cite{zeltner2020specular}, albeit at the cost of high variance.}
	Although it is possible to eventually find all solutions, the variance \note{could be arbitrarily high} \cite{fan23mpg}. Instead, we adopt stochastic sampling of regions (triangle tuples), effectively controlling the variance \note{introduced by random sampling.}
	
	\note{\paragraph{Analyzing the contribution of glossy triangles}
		For single scattering, \citet{Loubet20} derive a closed-form approximation of a glossy triangle's contribution. }They compute a position bound for each triangle using samples, which is not conservative, so it relies on path tracing to keep unbiasedness, producing visible fireflies. Even worse, for pure specular surfaces, path tracing cannot find caustics \note{at all} so it leads to bias. This inspires us to seek a theoretically conservative bound, \note{leveraging the Bernstein polynomials.}

	%
	%

	\paragraph{Beam tracing for caustics rendering}
	
	Our method revisits the concept of beam tracing \cite{Heckbert84}, traditionally applied to caustics rendering approximately. \citet{Watt90} pioneered \note{this manner} by generating a beam of light for each specular polygon and projecting it onto a receiver, defining its irradiance as proportional to the area ratio of these polygons. Despite subsequent improvements in accuracy \cite{Shinya87} and performance \cite{Iwasaki03}, they rely on heuristics that assume constant or linear variations in directions or irradiance. Even more, actual caustics do not necessarily form polygons on the receiver. These limitations motivate our more accurate modeling of caustic position and irradiance based on a rational formulation. 
	Besides, we only use precomputation to drive stochastic sampling in rendering. 
	
	\paragraph{Bernstein polynomials}
	
	The Bernstein polynomial basis \cite{Ber12} has been widely adopted in computer-aided design to model parametric curves and surfaces  \cite{Farouki2012TheBP}. A significant advantage is that the Bernstein coefficients of a polynomial offer valuable insights into its behavior over a finite interval, leading to numerous beneficial properties. In this work, we leverage a key property of the Bernstein coefficients, which provide conservative bounds for rational functions, demonstrating better tightness (see Fig. \ref{fig_interval}) than conventional methods like interval arithmetic \cite{Garloff2012BoundsOT}.

	\begin{figure}[t]
		\centering
		\begin{minipage}{\linewidth}
			\centering
			\includegraphics[width=0.95\linewidth]{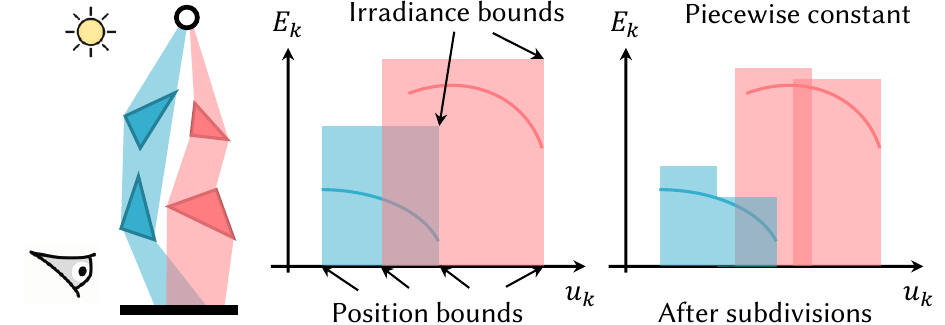}
		\end{minipage}
		\caption{
			\textbf{Overview of the precomputation pass.} We trace light beams passing through each triangle tuple and leverage their position and irradiance bounds to sample these triangle tuples. Note that our discussion of irradiance is primarily about its upper bound only.
		}
		\label{fig_overview}
	\end{figure}

	\section{Motivation and Overview}
	
	
	%
	
	Monte Carlo rendering of caustics involves sampling specular chains $\Bar{\boldsymbol{x}}$ comprised of specular vertices $\boldsymbol{x}_{1}, \boldsymbol{x}_{2}, \dots, \boldsymbol{x}_{k-1}$, which connects two non-specular endpoints $\mbx_0$ and $\mbx_k$. Typically, $\boldsymbol{x}_{0}$ lies on the light source, and $\boldsymbol{x}_{k}$ is a non-specular shading point. Each specular vertex $\boldsymbol{x}_{i}$ lies on a triangle\footnote{We first assume these triangles are given and defer how to incrementally find $\mathcal{T}_j$ given $\mathcal{T}_1,...,\mathcal{T}_{j-1}$ (for any $j>1$) to the implementation section.} $\mathcal{T}_{i}$. \note{We refer to previous works for a more gentle introduction of how this process works with a path tracer \cite{zeltner2020specular,fan23mpg,hanika2015manifold}.}
	
	Our method divides this sampling process into two \note{separate steps}: first, sampling some tuples of triangles given endpoints, and second, solving for all specular chains within each selected tuple $\mathcal{T}=(\mathcal{T}_1,...,\mathcal{T}_{k-1})$. 
	\note{Note that we focus on a single light source. If multiple emitters are involved, they should be treated as part of the tuple $\mathcal{T}$.} While existing works proposed ways to solve for paths within $\mathcal{T}$ \cite{wang2020path, walterSingleScatteringRefractive2009, fan24} (the second step), the sampling of $\mathcal{T}$ (the first step) is a clear gap.

	\paragraph{The goal of our method}
	
	We aim to sample a set of triangle tuples given endpoints according to their contributions. This requires a probability $P_\mathcal{T}$ (conditioned on $\mbx_k$) describing the chance we sample $\mathcal{T}$. 
	We compute this probability in a precomputation pass, which is stored on \note{non-specular} surfaces for later use in the rendering pass. 
	Precomputation takes the light source, specular caustics casters, and \note{non-specular} receivers as input but is independent of the camera. Instead of fitting from point samples that cannot preserve all information over the whole domain, we theoretically analyze the energy distribution of specular paths over finite intervals.

	\begin{figure}[t]
		\centering
		\begin{minipage}{\linewidth}
			\includegraphics[width=1\linewidth]{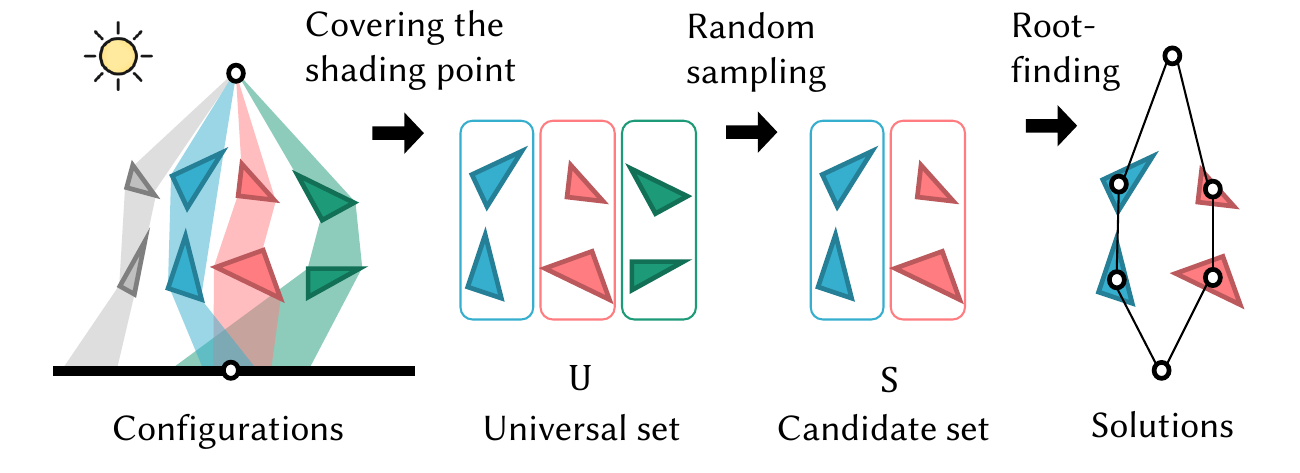}
		\end{minipage}
		\caption{
			\textbf{Overview of the rendering pass.} For a given shading point, the triangle tuples whose position bound covers it form a set $\mathrm{U}$. We aim to sample a subset $\mathrm{S} \subseteq \mathrm{U}$. For each triangle tuple $\ct \in \mathrm{S}$, we run existing root-finding methods to solve for admissible paths within it.
		}
		\label{fig_sample}
	\end{figure}
	
	\paragraph{Overall pipeline}
	
	In the precomputation pass, we trace beams from the light source passing through each tuple of specular triangles. This process is illustrated in Fig. \ref{fig_overview}. 
	For simplicity, we firstly consider point light sources only and a specific type of specular chain, that is, the chain length and the scattering type (reflection/refraction) at each vertex.
	This allows us to parameterize paths via the barycentric coordinates $\mbu_1$ of the first specular vertex $\mbx_1$.
	Each path intersects with the receiver $\mathcal{T}_k$ at vertex $\mbx_k$ with barycentric coordinates $\mbu_k$, which we formulate as a rational function of $\mbu_1$.
	
	Given a box on the domain of $\mbu_1$, the \textbf{position bound} describes the range of $\mbu_k$ where the corresponding paths hit the receiver, which indicates whether we need to consider a triangle tuple $\mathcal{T}$ at a shading point $\mbx_k$ on $\mathcal{T}_k$.
	However, the contribution of an admissible path could be small in some instances. Fortunately, we found that a rational function can bound the irradiance contribution. Therefore, we additionally introduce the \textbf{irradiance bound} that describes the range of irradiance received at $\mbx_k$. 
	These bounds are stored as (axis-aligned) boxes in the space of $\mbu_k$ with irradiance values.
	
	During the final rendering (Fig. \ref{fig_sample}), for a non-specular shading point (generated by, for example, path tracing), we query the boxes that cover this shading point and sample a subset of corresponding triangles according to their irradiance bound. We determine the sampling probability by optimizing an upper bound of variance.
	
	%
	%
	%
	
	All these bounds stem from the range bound of functions over a finite domain. Due to the availability of rational formulations, we obtain such bounds utilizing the Bernstein polynomials.


	
	\section{Bounds for specular paths}
	
	In this section, we aim to acquire a conservative bound of the irradiance distribution of caustics. We first introduce a property of Bernstein polynomials \cite{Farouki2012TheBP}, which enables a piecewise constant approximation of any rational function.

	\subsection{Bernstein bound for rational functions}
	
	\note{\textbf{The Bernstein basis}} for multivariate functions $p(\mbx)$ over $\mbx = (x_1, ..., x_m) \in \mathcal{U}^m = [0,1]^m$ are polynomials defined as
	\begin{equation}
		B_{\mbi,\mbn}(\mbx) = \prod_{j=1}^{m} 
		\binom{n_j}{i_j}
		x_j^{i_j} (1-x_j)^{n_j - i_j}
		,
	\end{equation}
	where $\binom{n_j}{i_j}$ is the binomial coefficient,
	$\mbn = (n_1, ..., n_m)$ is the degree of the polynomial, $\mbi = (i_1, ..., i_m)$ is the index of basis, and $j$ denotes the index of dimension\note{, for} all $j \in [1, m]$, \( i_j \in [0, n_j] \). 
	An essential property of Bernstein polynomials is that we can bound the range using its coefficients, which also generalizes to rational functions. 
	Specifically, consider a rational function \( f(\mbx) = \frac{p(\mbx)}{q(\mbx)} \) expressed in Bernstein coefficients \( b_{\mbi}(p) \) and \( b_{\mbi}(q) \) in the same degree:
	\begin{equation}
		p(\mbx) = \sum_{i_1=0}^{n_1} \dots \sum_{i_m=0}^{n_m} b_{\mbi}(p) B_{\mbi,\mbn}(\mbx)
	\end{equation}
	and similar for $q(\mbx)$. 
	Assuming that all \( b_i(q) \) have the same sign and are non-zero, the minimal and maximal coefficient ratios can bound\footnote{\note{
			To explain the high-level intuition that why Bernstein bounds are tighter than interval counterparts, we regard interval arithmetic as always using independent variables for different operands. For instance, suppose we are computing the bound of $f(x)=g(x) h(x)$ in $[0,1]$, with $g(x)=x$ and $h(x) = 4-x$. Using interval arithmetic, we obtain $[0,1] \times [3,4] = [0,4]$. This is because it completely ignores the correlation, effectively treating it as $x (4-y)$ with two independent variables $x$ and $y$. Instead, polynomial operations keep correlations due to the common variable $x$. As we obtain Bernstein coefficients $b_0=0, b_1=2, b_2=3$, we get a tighter bound $[0,3]$.}
	} the range of the function \note{\cite{Narkawicz12}}:
	\begin{equation}
		\min_{i_1=0}^{n_1} \dots \min_{i_m=0}^{n_m}
		\frac{b_{\mbi}(p)}{b_{\mbi}(q)} = \underline{f} \leq f(\mbx) \leq \overline{f} =
		\max_{i_1=0}^{n_1} \dots \max_{i_m=0}^{n_m}
		\frac{b_{\mbi}(p)}{b_{\mbi}(q)}.
		\label{eq_bbound}
	\end{equation}
	

	The above $f(x)$ corresponds to the barycentric coordinates of the receiver vertex $\mbu_k(\mbu_1)$ and the irradiance $E_k(\mbu_1)$ received there; we will present their detailed formulation in Secs. 4.3 and 4.4.
	In these cases, the restrictions on the denominator $q$ are not always satisfied. When not all $b_{\mbi}(q)$ have the same sign, we consider the \textbf{reciprocal} of $f(\mbx)$ so that if all $b_{\mbi}(p)$ have the same sign, we can still get a valid bound, though may be separated into two intervals. If both are unsatisfied, the result bound is the universal set $(-\infty, +\infty)$\note{, which typically occurs near grazing angles and focal points.}
	
	A constant bound with substantial internal variation is intrinsically loose. Fortunately, \note{the Bernstein bound of polynomials enjoys quadratic convergence with respect to the length of the interval \cite{Garloff1986}, so we can also improve the bound of rational functions via subdivisions \cite{Narkawicz12}}. This motivates us to use a \textbf{piecewise constant bound} of the function, which acts as a piecewise constant approximation of the function. The number of pieces controls the balance between tightness and computational cost. As the number of pieces tends to infinity, the bound converges to the actual value of the functions. \note{Thus, the infinite bound is a small portion after proper subdivisions, as shown in Figs. \ref{fig_teaser} and \ref{fig_eqtime}.}

	\subsection{Bounds in specular light transport}
	
	\begin{table}[tbp]
		\small
		\centering
		\caption{List of important symbols. }
		\label{tab:parameters}
		\begin{tabular}{ll}
			\toprule
			\textbf{Symbol} & \textbf{Description} \\ 
			\midrule
			$\ct_k$ & Receiver triangle \\
			$\ct = (\ct_1, ..., \ct_{k-1})$ & Specular triangle tuple \\
			\midrule
			$\boldsymbol{x}_{0}$ & Position of the point light \\ 
			$\overline{\mbx} = (\boldsymbol{x}_{1},\dots, \boldsymbol{x}_{k-1})$ & Position of the specular vertices \\
			$\boldsymbol{x}_{k}$ & Position of the shading point \\ 
			$\mbn_i$ & Un-normalized interpolated normal of $\mbx_i$ \\ 
			$\mbd_i$ & Position difference of vertices $\mbx_{i+1}$ and $\mbx_{i}$ \\ 
			$\mbu_i=(u_i, v_i)$ & Barycentric coordinate of  $\mbx_i$\\ 
			\midrule
			$\mbp_{i,0}, \mbp_{i,1}, \mbp_{i,2}$ & Vertex positions \\ 
			$\mbe_{i,1}, \mbe_{i,2}$ & Vector of triangle edges \\ 
			$\mbn_{i,0}, \mbn_{i,1}, \mbn_{i,2}$ & Vertex normals \\ 
			\midrule
			$E_{k}(\mbu_1)$ & Path's irradiance received at $\mbx_{k}$ \\
			$E(\ct, \mbu_k, \ct_k)$ & Tuple's irradiance received at $\mbx_{k}$ \\
			$\underline{f}, \overline{f}$ & Lower/upper bound of function $f$'s range \\
			\bottomrule
		\end{tabular}
		\label{tab_sym}
	\end{table}
	
	Now, we apply the bounding property of Bernstein polynomials to specular light transport. Given a triangle tuple \(\mathcal{T}\) and a box \(\bm{U}_1 = [\underline{u}_1, \overline{u}_1] \times [\underline{v}_1, \overline{v}_1]\), we define the \textbf{position bound} \(\bm{U}_k = [\underline{u}_k, \overline{u}_k] \times [\underline{v}_k, \overline{v}_k]\) as a box covering the range of \(\bm{u}_k(\bm{u}_1)\) on the receiver $\mathcal{T}_k$:
	\begin{equation}
		\underline{u}_k \le \inf_{\bm{u}_1 \in \bm{U}_1} u_k(\bm{u}_1), \quad
		\overline{u}_k \ge \sup_{\bm{u}_1 \in \bm{U}_1} u_k(\bm{u}_1).
	\end{equation}
	The same applies to \(\underline{v}_k\) and \(\overline{v}_k\). Similarly, the \textbf{irradiance bound} is the interval \(\boldsymbol{E}_k =  [\underline{E}_k, \overline{E}_k]\) that covers the range of \(E_k(\bm{u}_1)\). Note that all these quantities depend on \(\mathcal{T}\) and \(\mathcal{T}_k\); we omit them for simplicity. We summarize important symbols in Table \ref{tab_sym}. Note that throughout this paper, we use $\underline f$ and $\overline f$ for the bound we compute. It differs from the supremum, infimum, and range of a function.

	We compute these bounds leveraging the aforementioned ratios of  Bernstein coefficients. For instance,
	\begin{equation}
		\underline{u}_k = \min_{i_1=0}^{n_1} \dots \min_{i_m=0}^{n_m}
		\frac{b_{\mbi}^{u}(p)}{b_{\mbi}^{u}(q)},
		\quad
		\overline{u}_k = \max_{i_1=0}^{n_1} \dots \max_{i_m=0}^{n_m}
		\frac{b_{\mbi}^{u}(p)}{b_{\mbi}^{u}(q)}.
	\end{equation}
	Here, $b_{\mbi}^{u}$ refers to the Bernstein coefficients of $u_k(\mbu_1)$. 
	Likewise, $\underline{v}_k, \overline{v}_k$ and $\underline{E}_k, \overline{E}_k$ are computed from the Bernstein coefficients of $v_k(\mbu_1)$ and $E_k(\mbu_1)$, respectively.
	Nevertheless, the bounding property only works for rational functions, which requires formulating coordinates $\mbu_k(\mbu_1)$ and irradiance $E_k(\mbu_1)$ into rational functions while keeping the bound valid.

	\begin{figure}[t]
		\centering
		\begin{minipage}{\linewidth}
			\includegraphics[width=1\linewidth, trim=0 0 27 0]{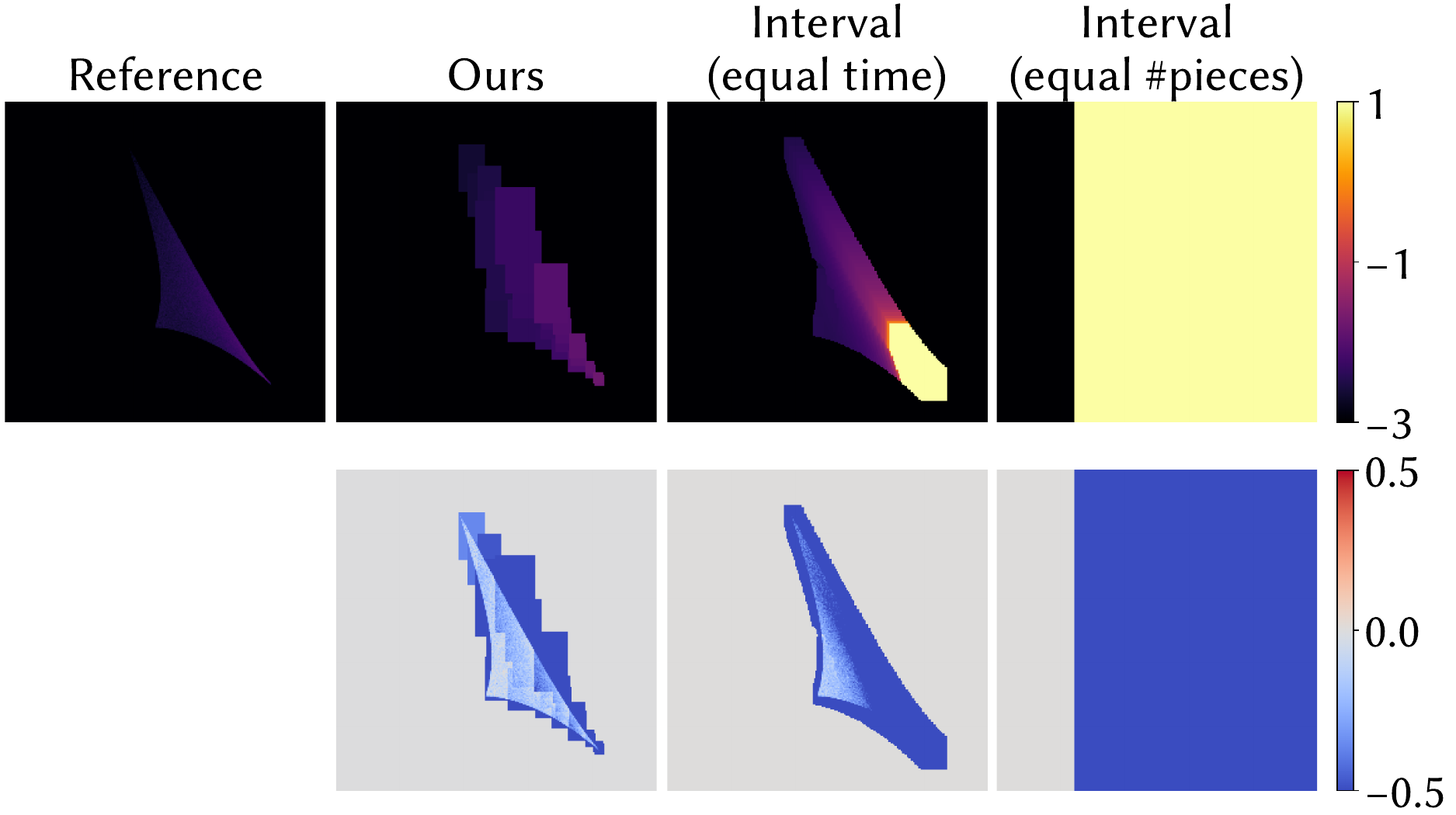}
			\includegraphics[width=1\linewidth]{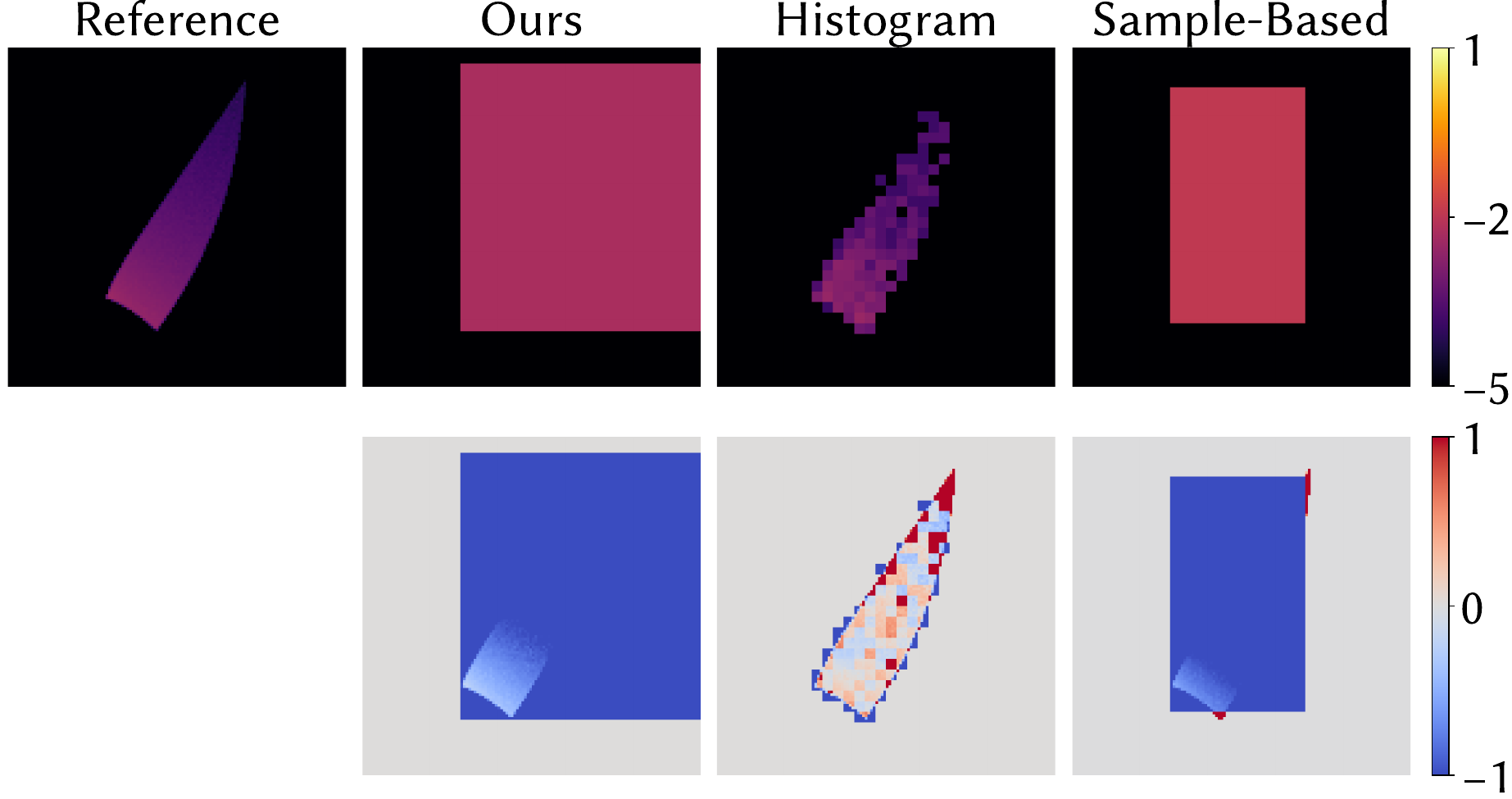}
		\end{minipage}
		\caption{
			\textbf{Visualizing the bound of caustics} cast by a single triangle reflector. We show the bound and its ratio with reference in the base 10 logarithmic space. \note{Note that the axes are barycentric coordinates $u_k$ and $v_k$ on the \textbf{receiver}. Thus, the position bounds of different pieces may overlap.}
			\textbf{Top:}
			Our bound is tighter than interval arithmetics \cite{wang2020path} in both an equal number of pieces and equal time.  Ours computes 50 pieces while interval arithmetic computes 4K pieces in equal time.
			\textbf{Bottom:}
			Utilizing 500 uniformly sampled paths (in roughly equal time) to fit irradiance distributions (Histogram) like path guiding \cite{jensenImportanceDrivenPath1995} or use the range of samples as bounds (Sample-Based) \cite{Loubet20} are not conservative and could result in fireflies or bias in rendering.
		}
		\label{fig_interval}
	\end{figure}

	\subsection{Rational formulation of vertex positions }
	\label{sec_vp}
	
	Our rational formulation of vertex positions (coordinates) builds upon rational coordinate mapping between vertices in a specular chain \cite{fan24}. Here, we briefly review the necessary formulas and refer to the original paper for derivation.
	
	\paragraph{Rational coordinate mapping}
	Formally, we use the barycentric coordinates $\mbu_i=\left(u_i,v_i\right)^\top$ to represent the vertex positions, normals, and differences between consecutive vertices:
	\begin{equation}
		\mbx_i = \mbp_{i,0} + u_i(\mbp_{i,1}-\mbp_{i,0}) + v_i (\mbp_{i,2}-\mbp_{i,0}),
		\label{eq_xi}
	\end{equation}
	\begin{equation}
		{\mbn}_i = \mbn_{i,0} + u_i (\mbn_{i,1}-\mbn_{i,0}) + v_i (\mbn_{i,2}-\mbn_{i,0}),
		\label{eq_ni}
	\end{equation} 
	\begin{equation}
		\mbd_i= \mbx_{i+1} - \mbx_i.
	\end{equation}
	Note that $\mbn_i$ and $\mbd_i$ are un-normalized. All these expressions are functions on  $\mbu_1$. For simplicity, we omit the function's variables.
	Each vertex can be represented by its preceding vertex's coordinates \cite{mollerFastMinimumStorage1997}:
	\begin{equation}
		\mbu_{i+1} =\left( \frac{
			(
			\tilde{\mbd}_i
			\times
			\mbe_{i+1, 2}
			)\cdot
			(\mbx_i - \mbp_{i+1, 0})
		}
		{
			(
			\tilde{\mbd}_i
			\times
			\mbe_{i+1, 2}
			)\cdot
			\mbe_{i+1, 1}
		},
		\frac{(
			(\mbx_i - \mbp_{i+1, 0})
			\times
			\mbe_{i+1, 1}
			)\cdot
			\tilde{\mbd}_i
		}
		{
			(
			\tilde{\mbd}_i
			\times
			\mbe_{i+1, 2}
			)\cdot
			\mbe_{i+1, 1}
		}\right)^\top.
		\label{eq_rct}
	\end{equation}
	Here, $\tilde{\mbd}_i$ is a scaled version of ${\mbd}_i$, which is determined by the scattering type at $\mbx_i$. We define $\mbe_{i,1} = \mbp_{i,1} - \mbp_{i,0}$ and similar for $\mbe_{i,2}$. The (un-normalized) outgoing direction at $\mbx_i$ is
	\begin{equation}
		\tilde{\mbd}_{i} = 
		\begin{cases}
			({\mbn}_i \cdot {\mbn}_i) \mbd_{i-1} -2(\mbd_{i-1} \cdot {\mbn}_i)
			
			{\mbn}_i , & \textrm{for reflection}
			\\
			\eta_i' \left(({\mbn}_i \cdot {\mbn}_i) \mbd_{i-1}  - (\mbd_{i-1} \cdot \mbn_i) \mbn_i\right) - \sqrt{
				\beta_i
			}
			\mbn_i, & \textrm{for refraction}
		\end{cases}
		\label{eq_refr_rational}
	\end{equation}
	with $\eta_i'$ being the ratio of index of refractions and
	\begin{equation}
		\beta_i
		= ( 1 - \eta_i'^2)
		({\mbn}_i \cdot {\mbn}_i) ({\mbd}_{i-1} \cdot {\mbd}_{i-1})
		+ \eta_i'^2		(\mbd_{i-1} \cdot \mbn_i)^2.
	\end{equation}
	The \note{square} root is not rational. Specular polynomials handle it by introducing a precomputed piecewise rational approximant \cite{fan24}. 
	However, this approximation introduces error and thus could lead to an invalid (i.e., not conservative) bound.
	
	Fortunately, we can correct it through auxiliary \textbf{remainder variables} that model the approximation error.
	
	\begin{figure}[t]
		\centering
		\begin{minipage}{\linewidth}
			\centering
			\includegraphics[width=0.9\linewidth]{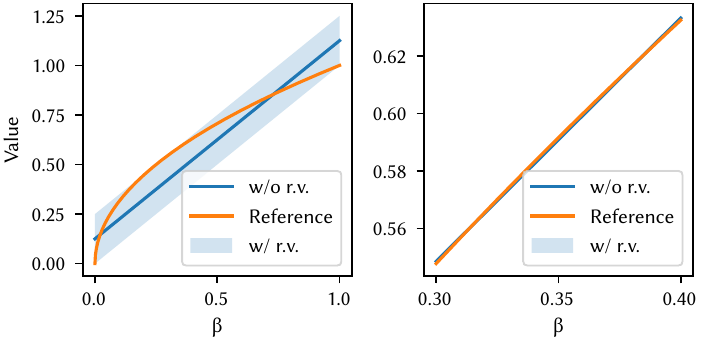}
		\end{minipage}
		\caption{
			\textbf{With remainder variables}, we correctly bound the range of $\sqrt{\beta}$, though looser than the proper range (left). Note that the actual range of $\beta$ is smaller, so the bound could be tighter (right).
		}
		\label{fig_remainder}
	\end{figure}

	\paragraph{Remainder variables}

	We introduce a remainder variable $\xi_i$ for each refractive vertex to compensate for the difference between $\sqrt {\beta_i}$ and a rational approximation $r(\beta_i)$. Now we can safely replace the original occurrence of $\sqrt {\beta_i}$ with a rational function
	\begin{equation}
		\tilde{r}(\beta_i, \xi_i) = r(\beta_i) + (1 - \xi_i) \underline \delta_i + \xi_i \overline \delta_i,
	\end{equation}
	where $\underline \delta_i$ and $\overline \delta_i$ is the bound of the approximation error
	$
	\delta_i(\mbu_i) = \sqrt {\beta_i}  - r(\beta_i).
	$
	Note that $\xi_i$ is independent of $\beta_i$.
	This introduces an extra dimension and corrects the bound, though it may become loose. 
	
	To guarantee the bound will converge to the true value as the domain of $\mbu_1$ gets infinitely small, we generate the approximation $r(\beta_i)$ on the fly, leveraging the range bound of $\beta_i$ on the current domain. Specifically, we use a linear approximation (Fig. \ref{fig_remainder}):
	\begin{equation}
		r(\beta_i) = a \beta_i + b.
	\end{equation}
	Please refer to the supplemental document for the proof and the closed-form calculation of parameters $a,b$, $\underline \delta_i$, and $\overline \delta_i$.

	Until now, by substituting Eq. \eqref{eq_rct} into Eq. \eqref{eq_bbound}, it is already possible to obtain a conservative position bound of specular vertices. Nevertheless, efficient sampling of triangles also requires additional information on their irradiance $E_k$.
	

	\subsection{Rational formulation of irradiance}
	
	\paragraph{Motivation}
	
	Analyzing irradiance enables us to sample triangles based on their contributions. It is particularly beneficial when specific triangles contribute minimal energy. After all, a nearly zero contribution is effectively equivalent to having no solution. An irradiance bound allows us to assign them a correspondingly low probability and focus more on paths with high contributions.
	
	\paragraph{Overview}

\begin{figure}[t]
		\centering
		\begin{minipage}{\linewidth}
			\centering
			\includegraphics[width=0.5\linewidth]{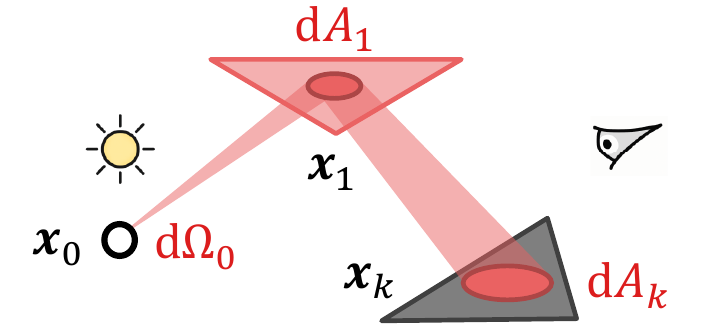}
		\end{minipage}
		\caption{
			\textbf{Illustration of the generalized geometric term (GGT)}. Light emitted in the differential solid angle $\mathrm{d}\Omega_0$ passing through specular surfaces finally hits the differential area $\mathrm{d}A_k$  on the receiver. Note that we use $\mathrm{d}\Omega_0$ instead of the projected solid angle $\mathrm{d}\Omega_0^{\perp}$ because we consider point light sources that emit intensity uniformly in different directions. }
		\label{fig_setup}
	\end{figure}

	The irradiance carried by the path $\overline{\mbx}$ received at $\mbx_k$ comprises several independent factors \cite{wang2020path, walterSingleScatteringRefractive2009, jakob2012manifold}:
	\begin{equation}
		E_{k}({\mbu_1}) = E_{k}(\overline{\mbx})=I_0 \frac{\mathrm{d}\Omega_{0}}{\mathrm{d}A_{k}} V(\overline{\mbx}) F(\overline{\mbx}).
	\end{equation}
	We parameterize by $\mbu_1$ since $\overline{\mbx}$ is determined by $\mbu_1$ when considering a specific chain type.
	Here, $I_0$ is the emitter's intensity\footnote{Since $I_0$ is completely scene-dependent, we ignore it (i.e., simply assuming $I_0=1$) across all the irradiance visualizations.  }, $\mathrm{d} \Omega_{0}$ refers to the differential solid angle at the point light source $\mbx_0$, and $\mathrm{d} A_k$ the differential area at vertex $\mbx_k$. $V$ represents the visibility term, and $F$ is the product of Fresnel terms at each vertex. 
	$\frac{\mathrm{d}\Omega_{0}}{\mathrm{d}A_{k}}$ is the generalized geometric term (GGT).
	As illustrated in Fig. \ref{fig_setup}, it describes the solid angular measure of emitted flux concentrated into a unit area on the receiver.
	For simplicity, we ignore $F$ and $V$ in the discussion since they are no greater than $1$. Thus, the upper bound of irradiance is still correct.
	We expand $\frac{\mathrm{d}\Omega_{0}}{\mathrm{d}A_{k}}$ using the chain rule, \note{which is similar to previous works \cite{Kaplanyan2014}}:
	\begin{equation}
		\frac{\mathrm{d}\Omega_{0}}{\mathrm{d}A_{k}}=
		\frac{\mathrm{d}\Omega_{0}}{\mathrm{d}A_{1}}
		\left|\frac{\partial \mbx_1}{\partial \mbu_1} \right|
		\left|\frac{\partial \mbu_1}{\partial \mbu_k} \right|
		\left|\frac{\partial \mbu_k}{\partial \mbx_k} \right|.
		\label{eq_dwda}
	\end{equation}
	We obtain rational expressions of these four terms separately. The key part is the Jacobian $\left| \frac{\partial \mbu_{1}}{\partial \mbu_{k}}\right|$; we discuss the remaining parts in the supplemental document. Here, we provide two approaches to obtain rational forms of $\left| \frac{\partial \mbu_{1}}{\partial \mbu_{k}}\right|$.

	\paragraph{Explicit differentiation}
	
	We initiate by computing the reciprocal using the form invariance of first-order differentials:
	\begin{equation}
		\left| \frac{\partial \mbu_{1}}{\partial \mbu_{k}} \right|
		=
		\frac{1}{\left| \frac{\partial \mbu_{k}}{\partial \mbu_{1}} \right|}
		=
		\frac{1}{
			\begin{vmatrix}
				\frac{\partial u_{k}}{\partial u_{1}} & \frac{\partial u_{k}}{\partial v_{1}} \\
				\frac{\partial v_{k}}{\partial u_{1}} & \frac{\partial v_{k}}{\partial v_{1}}
			\end{vmatrix}
			\label{eq_irr_term3a}
		}.
	\end{equation}
	This only requires the forward derivatives of the position expressions derived in Section \ref{sec_vp}. After all, the coefficients of a polynomial's partial derivative are trivially a linear combination of the original polynomial's coefficients, which is easy to compute. We acquire the final irradiance bound by substituting Eq. \eqref{eq_dwda} into Eq. \eqref{eq_bbound}.
	Note that the remainder variables in the square root approximation only guarantee the correctness of the position bound, so we use a derivative-aware approximation, which compensates not only the primal values but also the derivatives to $\beta_i$; see the supplemental document for details.

\begin{figure}[t]
		\centering
		\begin{minipage}{\linewidth}
			\includegraphics[width=1\linewidth]{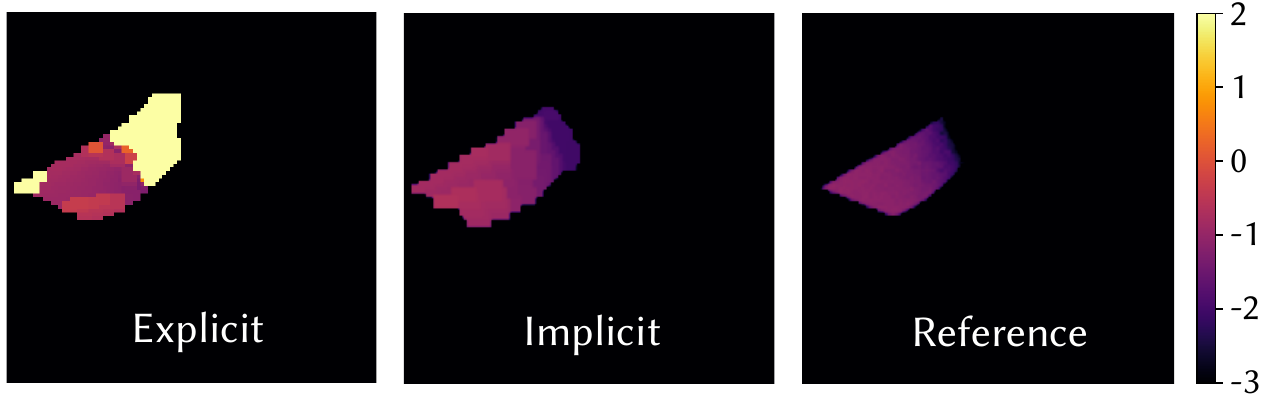}
		\end{minipage}
		\caption{
			The irradiance bound using explicit differentiation is extremely loose for cases involving nearly total internal reflection, even using $10^4$ pieces. Fortunately, the implicit differentiation succeeds. We show $\log_{10}(E_k)$. \note{Note that the axes are barycentric coordinates $u_k$ and $v_k$ on the \textbf{receiver}.}
		}
		\label{fig_refraction}
	\end{figure}

	Unfortunately, when the refracted angle is near $\pi/2$, $\beta_i$ tends to zero. Hence, the relative approximation error tends to infinity, so the bound becomes extremely loose (Fig. \ref{fig_refraction}). This motivates us to directly express $\left| \frac{\partial \mbu_{1}}{\partial \mbu_{k}} \right|$
	using the derivatives of implicit functions. 
	
	\paragraph{Implicit differentiation}

	Implicit differentiation does not require representing $\mbu_k$ as a rational function of $\mbu_1$, avoiding approximations on the refraction angle at a particular vertex $\mbx_k$.
	
	For refraction, we express the specular constraint on $\mbx_{k-1}$ as two polynomial functions $F$ and $G$ in variables $\mbu_1$ and $\mbu_k$. We use multivariate specular polynomials\footnote{The constraints $F$ and $G$ are polynomials in both $\mbu_1$ and $\mbu_k$ \cite{fan24}:
		\begin{equation}
			\begin{cases}
				F =\mbd_{k-1}^2 \left((\mbd_{k-2} \times \mbn_{k-1}) \cdot \mbb\right)^2 - \eta^2 \mbd_{k-2}^2 \left((\mbd_{k-1} \times \mbn_{k-1}) \cdot \mbb\right)^2,\\
				G = (\mbd_{k-2} \times \mbn_{k-1}) \cdot \mbd_{k-1}.
			\end{cases}
		\end{equation}
		Here, $\mbb$ represents an arbitrary vector with a non-zero norm, and $\eta$ is the ratio of the index of refraction. To avoid degenerated cases, we use the intersection of bounds computed using two different $\mbb$ vectors $\mbb_1 = (1, 0, 0)^\top$ and $\mbb_2 = (0, 1, 0)^\top$, respectively.} as $F$ and $G$.
	By solving
	\begin{equation}
		\begin{cases}
			\displaystyle
			F_{u_k} + F_{u_1} \frac{\partial u_1}{\partial u_k} + F_{v_1} \frac{\partial v_1}{\partial u_k} = 0, \\
			\displaystyle
			G_{u_k} + G_{u_1} \frac{\partial u_1}{\partial u_k} + G_{v_1} \frac{\partial v_1}{\partial u_k} = 0,
		\end{cases}
	\end{equation}
	we obtain
	\begin{equation}
		\frac{\partial u_1}{\partial u_k} = -\frac{ \frac{\partial F}{\partial u_k} \frac{\partial G}{\partial v_1} - \frac{\partial F}{\partial v_1} \frac{\partial G}{\partial u_k} }{ \frac{\partial F}{\partial u_1} \frac{\partial G}{\partial v_1} - \frac{\partial F}{\partial v_1} \frac{\partial G}{\partial u_1} }, \quad 
		\frac{\partial v_1}{\partial u_k} = \frac{ \frac{\partial F}{\partial u_k} \frac{\partial G}{\partial u_1} - \frac{\partial F}{\partial u_1} \frac{\partial G}{\partial u_k} }{ \frac{\partial F}{\partial u_1} \frac{\partial G}{\partial v_1} - \frac{\partial F}{\partial v_1} \frac{\partial G}{\partial u_1} }.
	\end{equation}
	The same goes for $v_k$. 
	The resulting Jacobian
	$\left| \frac{\partial \mbu_{1}}{\partial \mbu_{k}} \right| = \begin{vmatrix}
		\frac{\partial u_{1}}{\partial u_{k}} & \frac{\partial u_{1}}{\partial v_{k}} \\
		\frac{\partial v_{1}}{\partial u_{k}} & \frac{\partial v_{1}}{\partial v_{k}}
	\end{vmatrix}$
	includes both $\mbu_1$ and $\mbu_k$. Directly substituting $\mbu_k$ using an expression of $\mbu_1$ results in extremely high degrees. We instead treat $\mbu_1$ and $\mbu_k$ as independent variables within the position bound, but we still substitute $\mbu_2, ..., \mbu_{k-1}$ with rational functions in $\mbu_1$. \note{This step extends a 2D manifold to its 4D superset, so the bound may become looser but still valid.}
	Finally, we reach the irradiance bound by substituting Eq. \eqref{eq_dwda} into Eq. \eqref{eq_bbound}.
	\note{We present the degree of final expressions and complexity analysis in the supplemental document.}

	\begin{figure}[t]
		\centering
		\begin{minipage}{\linewidth}
			\includegraphics[width=1\linewidth]{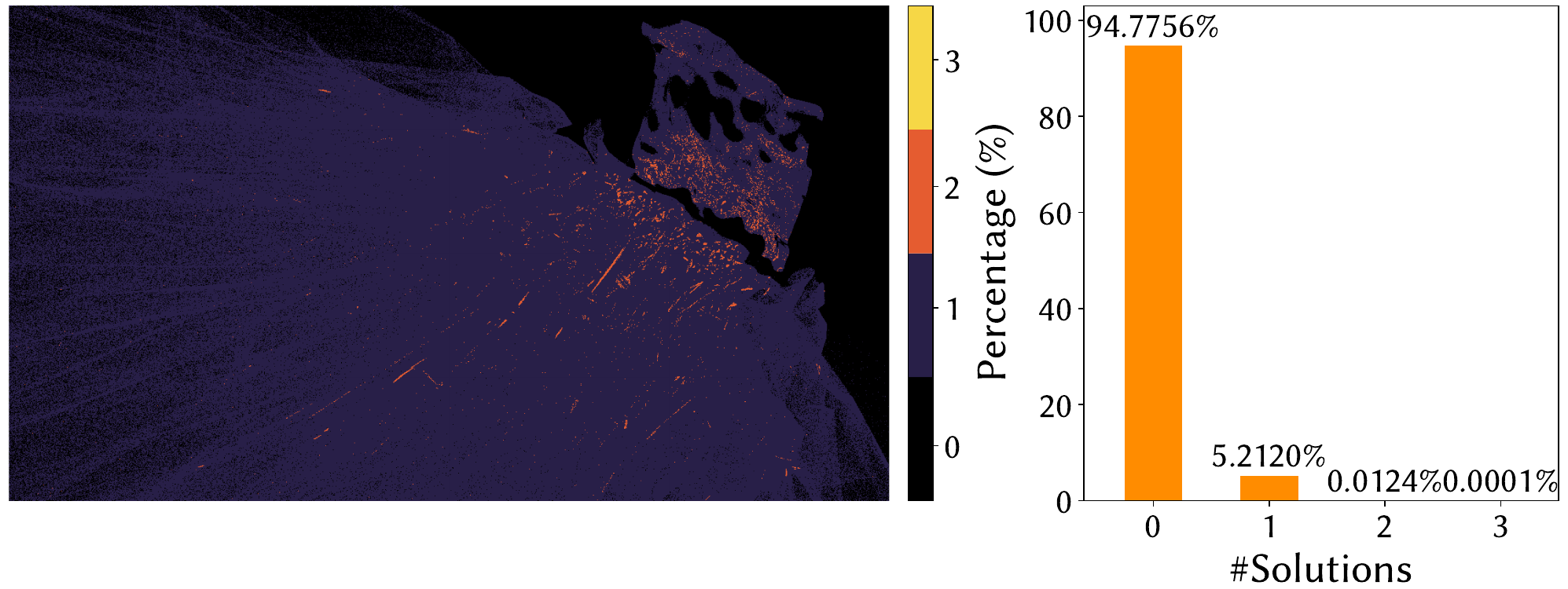}
		\end{minipage}
		\caption{
			\textbf{The number of solutions per tuple.}
			In the Plane scene, nearly all tuples have no more than $1$ solutions, so we set $m \le 1$ for rendering experiments. \textbf{Left}: the \textit{maximal} number of solutions per tuple for each shading point. \textbf{Right}: distributions of the number of solutions per tuple.}
		\label{fig_nroots}
	\end{figure}

	\subsection{Total irradiance contribution of a triangle tuple}
	
	The above-discussed irradiance $E_k(\mbu_1)$ pertains to an admissible specular path. Since we aim at sampling triangle tuples, we need to know the total irradiance contribution \(E(\mathcal{T}, \bm{u}_k, \mathcal{T}_k)\) of each tuple, a summation over all admissible paths' contribution within $\mathcal{T}$:
	\begin{equation}
		E(\mathcal{T},  \bm{u}_k, \mathcal{T}_{k}) =
		\sum_{\bm{u}_k = \bm{u}_k(\bm{u}_1)} E_k(\bm{u}_1).
		\label{eq_pertuple}
	\end{equation}
	\(E(\mathcal{T}, \bm{u}_k, \mathcal{T}_k)\) is no greater than $m$ times the path's irradiance bound. Here, $m$ denotes the number of solutions within $\ct$. For $\mbu_k$ not covered by the position bound, \(m\) must be zero. 
	For $\mbu_k$ covered by the position bound, following the common assumption that at most one solution exists when triangles are small, we set $m=1$ in our experiments\footnote{Even if \( m \) exceeds our assumption, it will not introduce bias. 	
		However, the variance may increase beyond expected levels. 
		$m$ is provably finite for triangle meshes \cite{wang2020path}.
		A theoretically strict upper bound of $m$ can be derived from the degree of specular polynomials, but is too loose in practice.}. \note{We show an example in Fig. \ref{fig_nroots}.}
	
\begin{figure*}[t]
		\centering
		\begin{minipage}{\linewidth}
			\includegraphics[width=1\linewidth, trim=0 10 0 10, clip]{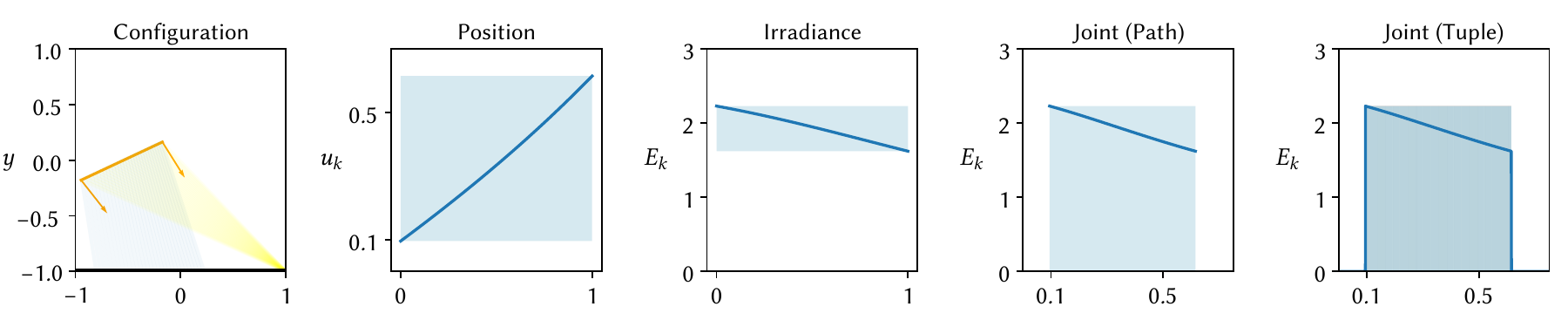}
			\includegraphics[width=1\linewidth, trim=0 8 0 26, clip]{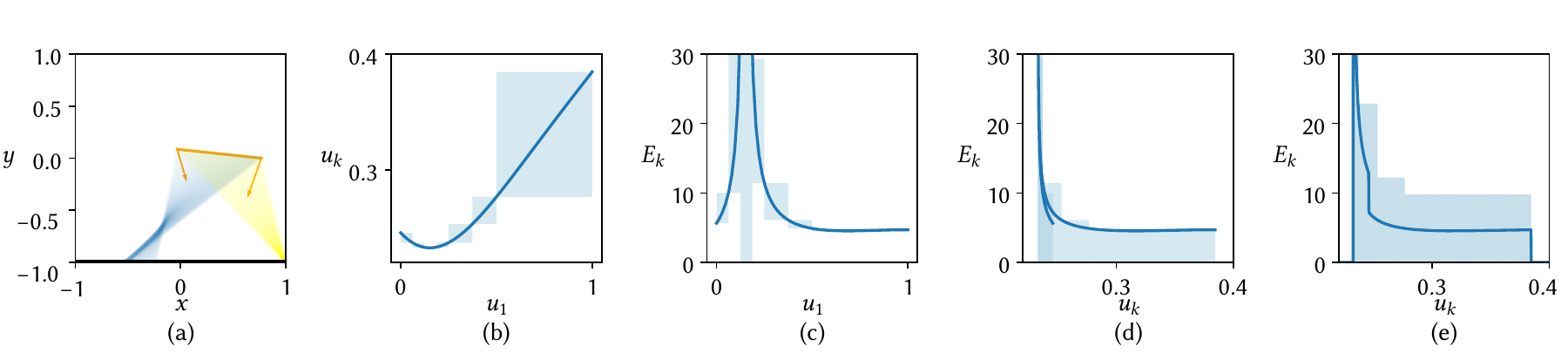}
		\end{minipage}
		\caption{
			\textbf{2D examples.}
			\textbf{(a)} The incident light (yellow) hits a specular triangle (orange, with interpolated normals). Reflected rays (blue) hit the receiver (black, bottom).
			\textbf{(b)} We show reference as solid curves and our bound as a shaded region. We perform subdivision on the domain of $u_1$ to acquire a piecewise constant bound.
			\textbf{(c)} At the singular point where $\frac{\mathrm{d}u_k}{\mathrm{d}u_1}=0$, the irradiance tends to infinity, so we cannot obtain a finite bound. 
			\textbf{(d)} A joint use of position and irradiance bound for bounding the \textbf{per-path} irradiance contribution at any position. Note the positional overlap and the irradiance singularity in the bottom example.
			\textbf{(e)} Eq. \eqref{eq_pertuple} and Eq. \eqref{eq_pertuple_b} provide the reference and bound of irradiance \textbf{per tuple}, respectively. Here, we set $m=1$ (top) and $m=2$ (bottom). \note{Note that $u_1$ is hidden by the transform from (b) and (c) to (d), thus it is \textbf{never} stored. The final storage for each $\mathcal{T}$ only involves $u_k$ and $E_k$.}
		}
		\label{fig_bound}
	\end{figure*}

	When considering subdivisions on the domain of $\mbu_1$, we should use the maximum irradiance bound among the pieces whose position bounds cover the shading point instead.
	Suppose the union of disjoint rectangles \(\bm{U}_1^1, \ldots, \bm{U}_1^n\) covers \(\mathcal{U}^2\). Each \(\bm{U}_1^i\) corresponds to a position bound \(\bm{U}_k^i\) and an irradiance bound \(\bm{E}_k^i\). 
	We can give an upper bound of the total irradiance contribution:
	\begin{equation}
		E(\mathcal{T},  \bm{u}_k, \mathcal{T}_{k})
		\le
		\tilde{E}(\mathcal{T},  \bm{u}_k, \mathcal{T}_{k})
		=
		m \max_{\bm{u}_k \in \bm{U}_k^i} \overline{E}_k^{i}.
		\label{eq_pertuple_b}
	\end{equation}
	\note{Note that the symbol $\overline E$ refers to per-path irradiance bound, while $\tilde E$ is per-tuple.}
	See proofs and detailed discussions in the supplemental document. Lastly, we showcase two 2D examples to demonstrate the overall process of the precomputation pass in Fig. \ref{fig_bound}.

	\section{Bound-driven sampling of triangle tuples}
	
	The previously computed bound provides substantial information about caustics, which we now leverage to sample triangle tuples effectively. 
	Our discussion in this section is under a specific shading point $\mbx_k$, which determines $\mathcal{T}_k$ and $\mbu_k$, so we omit these symbols for simplicity. For instance, $E(\ct)$ is the shorthand of $E(\ct, \mbu_k, \mathcal{T}_k)$, which represents the irradiance contributed by $\ct$ at $\mbx_k$.
	
	\subsection{Problem definition}

	
	There is a finite set $\mathrm{U}$ of different triangle tuples $\mathcal{T}$ whose position bounds cover the given shading point $\mbx_k$. Our goal is to estimate the sum of irradiance $E(\ct)$ over all triangle tuples $\ct \in \mathrm{U}$, i.e.,
	\begin{equation}
		E  = \sum_{\ct \in \mathrm{U}} E(\ct).
	\end{equation}
	For simplicity, we discuss the diffuse receiver, where the sum $E$ decides outgoing radiance $ L = \rho E$, with $\rho$ being the albedo. For glossy shading points, a product with the BSDF is required.
	
	We aim to sample a \textbf{candidate triangle tuple set} $\mathrm{S} \subseteq \mathrm{U}$. After that, we solve for admissible paths within each $\ct \in \mathrm{S}$ and sum their contributions (dividing the corresponding sampling probability). 
	Here, the key design choice is determining the number of tuples to sample and the probability with which each tuple is chosen.

	\subsection{Sampling a single triangle tuple}
	
	A straightforward way is to importance sample \cite{Veach98} triangle tuples proportional to their irradiance contributions.
	Theoretically, this is the optimal probability for a one-sample estimator. However, the true irradiance of each triangle tuple is \note{intractable}; we can only use the bound instead:
	\begin{equation}
		P_\textrm{1}(\ct) = \frac{\tilde E(\ct)} {\sum_{\ct' \in \mathrm{U}} \tilde E(\ct')}.
	\end{equation}
	Nevertheless, $\tilde E(\ct)$ is sometimes substantially higher than $E(\ct)$ due to a \note{loose} bound (either position or irradiance). This could lead to an unmanageable increase in variance.
	
	Practically, there is an intrinsic trade-off among validity, tightness, and computational cost of bounds. This motivates us to resort to a new category of estimators that possibly sample multiple tuples to guarantee the variance is controllable.

	\subsection{Sampling multiple triangle tuples}
	
	\begin{algorithm}[t]
		\small
		\caption{Estimator that samples multiple triangle tuples}
		\begin{algorithmic}[1]
			\REQUIRE Probabilities \( P_\ct \) for each triangle tuple \( \ct \in \mathrm{U} \)
			\ENSURE Selected tuple set \( \mathrm{S} \), estimated irradiance \( \langle E \rangle \)
			\STATE  \( \mathrm{S} \gets \emptyset \), \( \langle E \rangle \gets 0 \)
			\FOR{\( \ct \in \mathrm{U} \)}
			\STATE $r \gets \texttt{rng()}$
			\COMMENT{Generate a random number in $ [0,1]$} 
			\IF{\( r < P_\ct \)} 
			\STATE \( \mathrm{S} \gets \mathrm{S} \cup \{\ct\} \) \COMMENT{Select a triangle tuple $\ct$}
			\STATE \( 
			\displaystyle \langle E \rangle \gets \langle E \rangle + \frac{E(\ct)}{P_\ct} \) \COMMENT{Add the contribution of all solutions in $\ct$}
			\ENDIF
			\ENDFOR
			\RETURN \( \mathrm{S}, \langle E \rangle \)
		\end{algorithmic}
		\label{alg}
	\end{algorithm}

	
	Instead of considering which tuple is sampled, we allow multiple tuples to be selected and evaluate the decision to choose each tuple. This effectively unifies importance sampling and selective activation (i.e., deciding where to enable specular path sampling) \cite{Loubet20, fan23mpg} in the same framework. 
	
	Formally, we introduce $P_\ct$, which represents the discrete probability that $\ct$ is chosen, i.e., $P_\ct = \mathbb{P}[\ct \in \textrm{S}]$. Note that $P_\ct$ is not necessarily normalized, i.e., $\sum_{\ct \in \mathrm{U}} P_\ct \neq 1$. The estimator
	\begin{equation}
		\langle E \rangle = \sum_{\ct \in \mathrm{S}}  \frac{E(\ct)}{P_\ct}
	\end{equation}
	is unbiased as long as $P_\ct > 0$ for all $\ct$ that satisfies $E(\ct) > 0$. We summarize the sampling process in Algorithm \ref{alg}.
	
	The time complexity of Algorithm \ref{alg} is $\mathcal{O}(\vert \mathrm{U} \vert)$. Such a brute-force implementation already works since the root-finding process is complex.
	Yet, when $\mathrm{U}$ is substantially larger than $\mathrm{S}$, we can reduce the complexity to $\mathcal{O}(|\mathrm{S}| \log(|\mathrm{U}|))$.
	We pack tuples into groups (denoted as $\mathrm{B}$) whose sum of probabilities is no greater than one as a classical bin-packing problem. In each group, we simply importance sample one tuple. We use the first-fit greedy algorithm with $\mathcal{O}(|\mathrm{U}|)$ time preprocessing and a guaranteed approximation ratio of $2$. 
	
	
	As an important property, variance measures the quality of sampling. A significant benefit of our multi-sample estimator is that we can represent the upper bound of the (population) variance $\sigma^2$ (a.k.a. $\mathbb V [\langle E \rangle]$) only in terms of $\tilde E(\ct)$, $P_\ct$, and a constant $\mu$:
	\begin{equation}
		\sigma^2 = \mu_2 - \mu^2 \le \mu_2,  \quad
		\mu_2 \le \overline{\mu_2} = \sum_{\ct \in \note{\mathrm{U}}} \frac{{\tilde E^2}(\ct)}{P_\ct}.
		\label{eq_varbound_u}
	\end{equation}
	Here, $\mu$ denotes the mean value, and $\mu_2$ is the second-order moment.
	
	The irradiance bound ${\tilde E}(\ct)$ is precomputed and fixed now, so $\overline{\mu_2}$ is controlled only by $P_\ct$. A good design of $P_\ct$ should have a relatively low variance. Thus, we determine $P_\ct$ by optimizing $\overline{\mu_2}$.

	\subsection{Optimized sampling probabilities}
	
\begin{figure}[t]
		\centering
		\begin{minipage}{\linewidth}
			\includegraphics[width=1\linewidth, trim=200 0 0 0, clip]{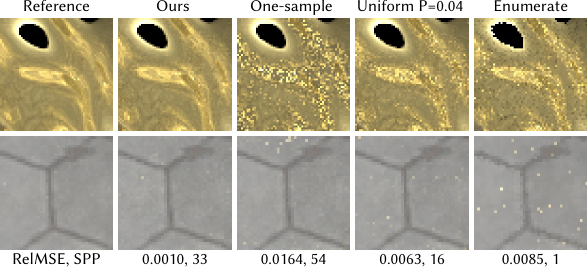}
		\end{minipage}
		\caption{\textbf{Validations on different sampling schemes.} While it is possible to use the position bound only to render a pixel-perfect image (Enumerate), it increases the runtime cost for each spp, leading to high variance in equal time comparison and obvious aliasing artifacts. Uniformly assigning $P$ introduces visible noise. Conventional importance sampling (One-sample) easily causes extremely high variance when the bound is loose. Our multi-sample estimator leveraging the irradiance bound performs the best.
		}
		\label{fig_cmp_sample}
	\end{figure}

	We aim to minimize \( \overline{\mu_2} \) given the expected number of candidates $W=\mathbb{E}[|\mathrm{S}|]$. Note that we first treat $W$ as a given parameter. Since $\mathrm{S} \subseteq \mathrm{U}$, we assume $W \le \left \vert \mathrm{U} \right \vert$.
	This is a nonlinear optimization with linear equality and inequality constraints:
	\begin{equation}
		\begin{aligned}
			\min_{P} \quad & \sum_{\ct \in \mathrm{U}} \frac{{\tilde E}^2(\ct)}{P_\ct}, \quad
			\\
			\textrm{s.t.} \quad &  \sum_{\ct \in \mathrm{U}} P_\ct = W,
			\\
			&  0 \le P_\ct \le 1, \quad \forall \ct \in \mathrm{U}.
		\end{aligned}
	\end{equation}
	According to the \note{Karush-Kuhn-Tucker (KKT)} conditions, we solve the above optimization using the Lagrangian with Lagrange multipliers $\lambda$ and $\lambda_\ct$:
	\begin{equation}
		\mathcal{L} = \sum_{\ct \in \mathrm{U}} \frac{{\tilde E}^2(\ct)}{P_\ct} 
		-
		\lambda \left(\sum_{\ct \in \mathrm{U}} P_\ct  - W\right)
		- \sum_{\ct \in \mathrm{U}} \lambda_\ct(P_\ct - 1)
	\end{equation}
	and obtain that for each $\ct$, either $P_\ct = 1$ or $P_\ct = \gamma {\tilde E}(\ct)$ is satisfied. Therefore, we reach the final probability
	\begin{equation}
		P_\ct = \min \left(\gamma{\tilde E}(\ct), 1\right)	
	\end{equation}
	with $\gamma$ being a constant parameter. Note that the above equation naturally satisfies $P_{\ct} \ge 0$, so we can ignore this constraint. In contrast, the condition $P_{\ct} \le 1$ must be considered explicitly as a key difference from continuous cases  \cite{rathVarianceawarePathGuiding2020}. \note{As a summary, we compare different estimators we discussed in Fig. \ref{fig_cmp_sample}.}
	
	Additionally, the impact of the parameter $\gamma$ is intuitive, as shown in Fig. \ref{fig_par1}.
	While it is possible to let users specify $\gamma$, obtaining it automatically from the expected number of samples $W$ or variance is also feasible since there is a one-to-one mapping between these parameters.
	See the supplemental document for details.

	\section{Evaluation}
	
	We implement our method in Mitsuba 0.6 \cite{Mitsuba}. The precomputation partially utilizes the Numba JIT compiler. All timings are conducted on an Intel Core i9-13900KF processor.
	
	\subsection{Implementation}
	
	The supplemental document provides detailed algorithms, complexity analysis, and pseudo-code snippets. We briefly outline some important design choices here.
	
	\paragraph{Tuple construction}
	
	In precomputation, we construct triangle tuples by extending triangles at the end of a given prefix. For $\mathcal{T}_1,...,\mathcal{T}_i$, we compute the bound of $\mbx_i$ and $\mbd_i$. Then, we traverse the bounding volume hierarchy (BVH) to find all possible $\mathcal{T}_{i+1}$ according to whether the bound of $(\mbx_{i+1} - \mbx_i) \times \mbd_i$ covers zero.

	\paragraph{Bound storage}
	
	We rasterize bounds into a grid parameterized by the texture coordinates. Each cell stores a list containing several pairs of irradiance bound $\tilde{E}(\mathcal{T},  \bm{u}_k, \mathcal{T}_{k})$ and triangle indices of $\mathcal{T}$. We splat the irradiance bound to each cell intersecting with the position bound. We choose uniform $512 \times 512$ grids for simplicity.
	\note{During rendering, for each shading point, we use its coordinate as $\mbu_k$ to query the grid on texture space, returning a list of tuples.}

	\begin{figure}[t]
		\centering
		\begin{minipage}{\linewidth}
			\includegraphics[width=1\linewidth]{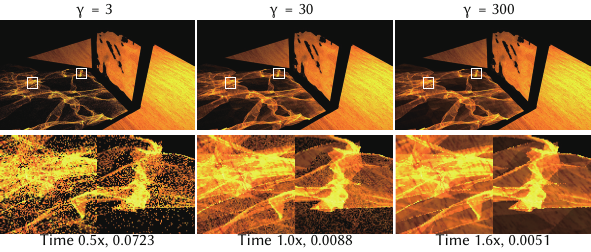}
		\end{minipage}
		\caption{ \textbf{The impact of the sampling parameter $\gamma$.} All methods use 4 spp for path tracing. \note{Higher gamma reduces variance by increasing sampling probability. Notably, the probability easily reaches $1$ (thus no variance) for high-energy tuples, while noise still persists for low-energy ones.}
		} 
		\label{fig_par1}
	\end{figure}
	
	\begin{figure}[b]
		\centering
		\begin{minipage}{\linewidth}
			\includegraphics[width=1\linewidth]{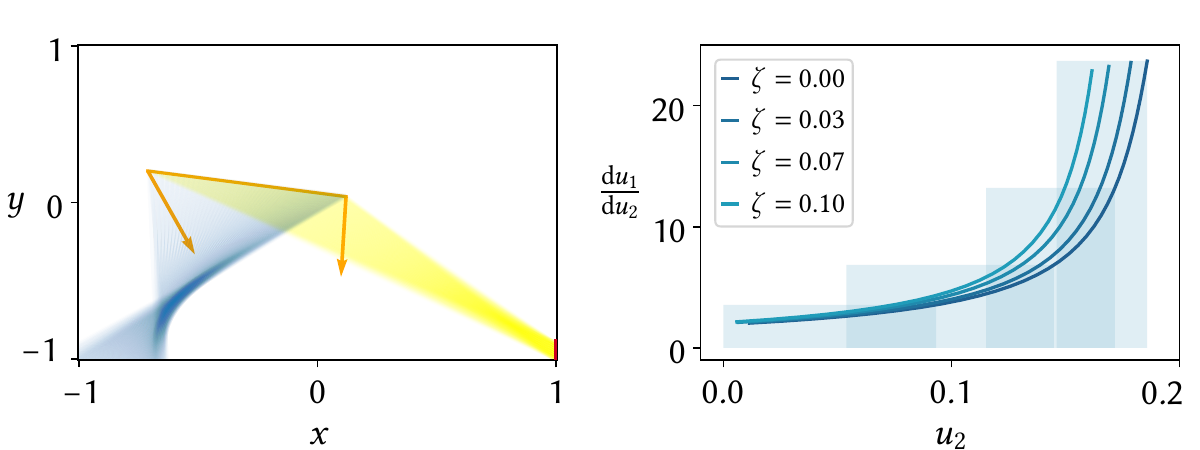}
		\end{minipage}
		\caption{
			\textbf{Handling area light sources.} We introduce an extra variable $\zeta$ to represent the position along the line light source. The rational functions \note{have} two variables $u_1$ and $\zeta$. We succeeded in bounding the irradiance distribution of all points on the area light source. The curves correspond to different positions (denoted as $\zeta$) on the light source. 
		}
		\label{fig_area}
	\end{figure}

	\begin{figure*}[t]
		\centering
		\begin{minipage}{\linewidth}
			\includegraphics[width=0.786\linewidth]{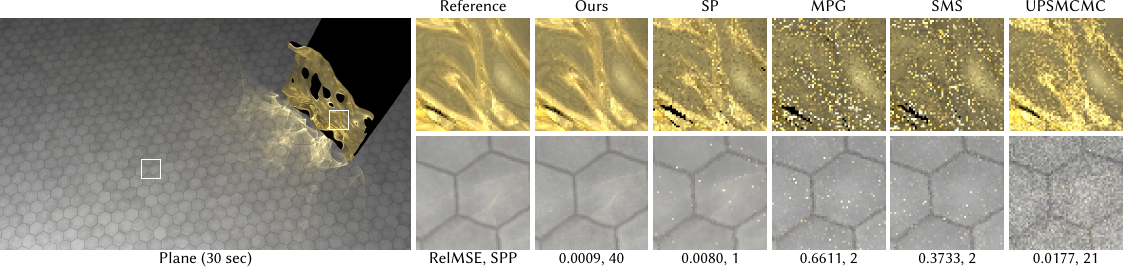}
			\includegraphics[width=0.211\linewidth]{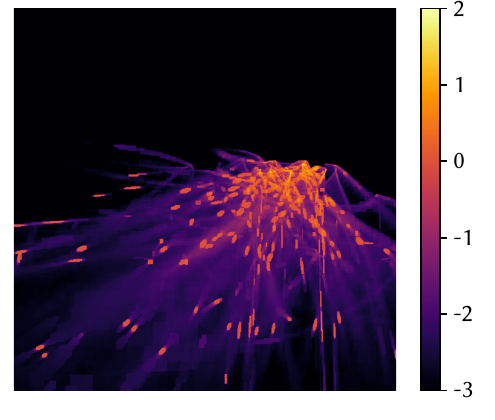}
			\includegraphics[width=0.786\linewidth]{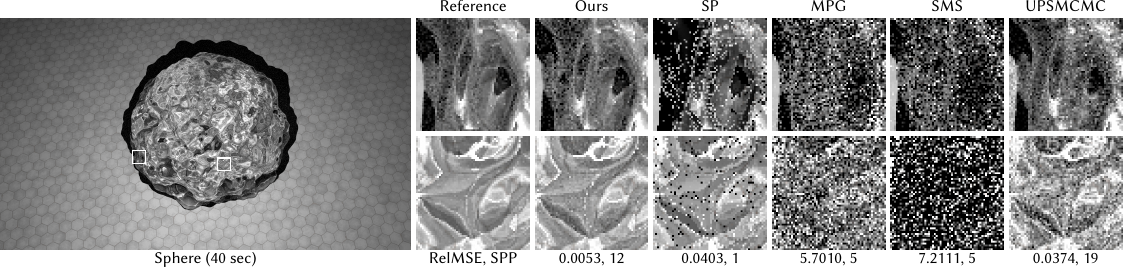}
			\includegraphics[width=0.211\linewidth]{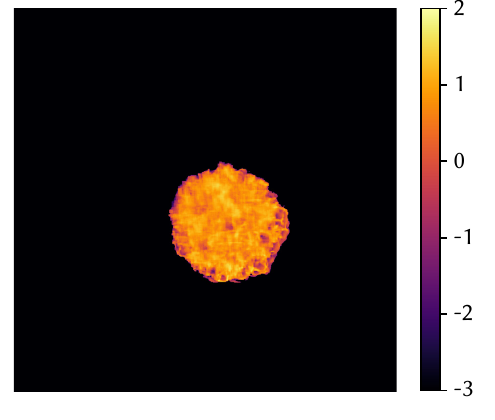}
		\end{minipage}
		\caption{
			\textbf{Equal-time comparisons on single scattering.} Precomputation takes 2.1 sec and 2.4 sec, respectively. We visualize irradiance bounds (in the base 10 logarithmic space) summed over tuples.
			We compare with Specular Polynomials (SP) \cite{fan24}, Manifold Path Guiding (MPG) \cite{fan23mpg}, Specular Manifold Sampling (SMS) \cite{zeltner2020specular}, and \note{Metropolised} Bidirectional Estimators (UPSMCMC) \cite{sikRobustLightTransport2016}.
		}
		\label{fig_eqtime}
	\end{figure*}

	\paragraph{Domain subdivision}
	
	Our bound is valid but loose when the triangles are large, leading to slow rendering. To compute the piecewise constant bound, we recursively subdivide the box domain of $\mbu_1$ into four boxes at the midpoint. The subdivision stops when
	\begin{itemize}
		\item the area of $\bm{U}_k$ is less than $\sigma = 10^{-4}$,
		\item  the ratio $\overline{E_k}/\underline{E_k}$ is smaller than a threshold\footnote{We set the approximation ratio $\alpha$ to 2 for single scattering and 10 for multiple ones.} $\alpha$, or
		\item the subdivision depth reaches a limit that varies among scenes.
	\end{itemize}
	Note that we do not subdivide $\mathcal{T}_2, \dots \mathcal{T}_k$, and the domain subdivision process is completely optional. 
	\note{Again, for different pieces after subdivisions, their domains (i.e., the range of $\mbu_1$) are completely disjoint, but position bounds on $\mbu_k$ may overlap with each other.}

	\paragraph{Domain initialization}
	
	For each triangle tuple, the effective domain of $\mbu_1$ is usually smaller than $\mathcal{U}^2$. Therefore, for each $\ct$, we first compute the bound of $\mbu_1$ where the corresponding $\mbu_k \in \mathcal{U}^2$, implemented using a recursive subdivision of at most 100 pieces.
	
	\paragraph{Degree reduction}
	
	We found the numerical stability and complexity become infeasible when the degree exceeds a certain threshold, e.g., 40. Thus, we convert these high-degree polynomials into low-degree ones and add a new remainder variable to maintain bounding validity. 
	Specifically, we fit low-degree approximants using linear regression based on singular-value decomposition. Each reduction also eliminates all existing remainder variables.

	\paragraph{Area light sources}
	
	Our method easily generalizes to area light sources by incorporating two additional variables in the expression for $\mbx_0$. In Fig. \ref{fig_area}, we present a 2D example for illustration.
	
	\paragraph{Root-finding}
	
	Rendering caustics requires finding admissible specular paths within the sampled tuple. We employ specular polynomials \cite{fan24} for single scattering.
	However, existing deterministic methods fail to find all solutions in multiple scattering while maintaining a low computational cost.
	Therefore, we choose Newton's method, assessing two different schemes of initialization and weighting to evaluate various aspects:
	
	\begin{itemize}
		
		\item \textbf{Deterministic} (Det) initialization \cite{wang2020path} may leak solutions but does not introduce additional variance, which helps evaluate the amount of variance introduced by our proposed triangle sampling. 
		
		\item \textbf{Stochastic} initialization with an unbiased\footnote{We briefly note that it is possible to combine a \textbf{stochastic} initialization with a \textbf{biased} weighting scheme \cite{zeltner2020specular}, which also suffers from energy loss but in a smoother pattern. These biased estimations remain below or equal to the ground truth, ensuring the overall second moment is still controllable.} weighting (Stoc) \cite{zeltner2020specular} helps validate the overall unbiasedness. However, it could introduce outliers. 
		
	\end{itemize}
	
	%

	\subsection{Equal-time comparisons}
	%
	%
	%
	%
	
\begin{figure*}[t]
		\centering
		\begin{minipage}{\linewidth}
			\includegraphics[width=1\linewidth]{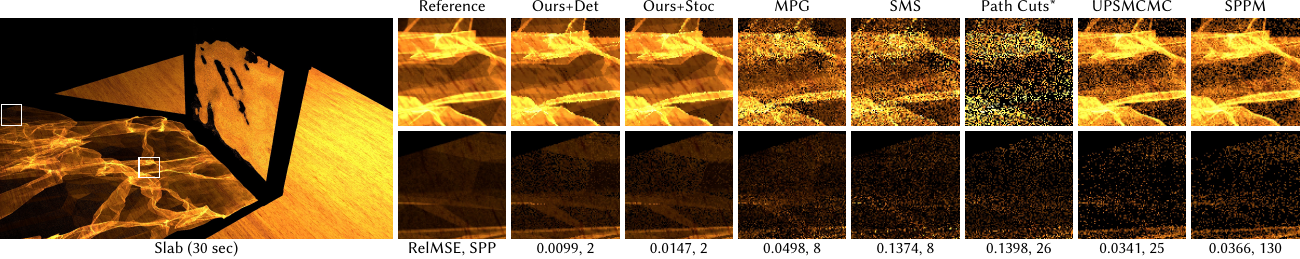}
			\includegraphics[width=1\linewidth]{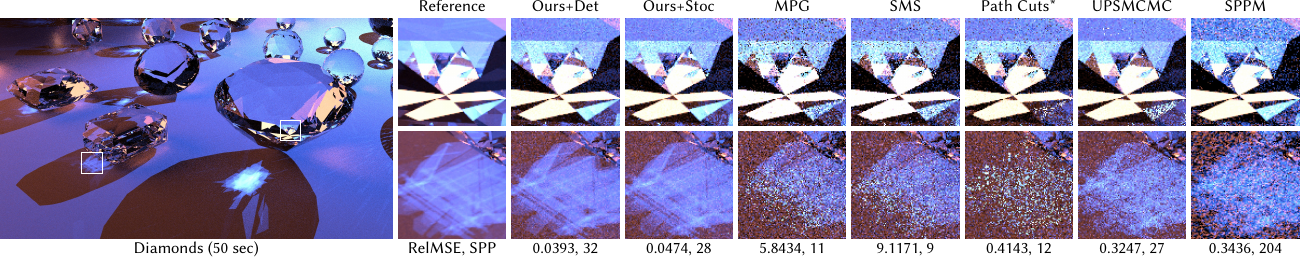}
		\end{minipage}
		\caption{
			\textbf{Equal-time comparisons on double scattering.} Precomputation time is included, which takes 23 sec and 25 sec, respectively. We combine our triangle sampling with deterministic (Det, biased) and stochastic (Stoc, unbiased) initialization for Newton's iteration-based root-finding. Since the original Path Cuts is extremely slow, we use a modified version (Path Cuts*) that samples 1\% paths, which already takes 60× more time than the other methods.
		}
		\label{fig_eqtime2}
	\end{figure*}

	In Figs. \ref{fig_eqtime} and \ref{fig_eqtime2}, we compare our method to several approaches:
	\begin{itemize}
		\item Deterministic search. We compare with specular polynomials \cite{fan24} for one bounce and Path Cuts \cite{wang2020path} for multiple bounces. Note that specular polynomial also uses Path Cut's interval tests to select triangles. For multiple scattering, we also note that the interval tests are extremely slow. To ensure a fair comparison with roughly equal time, we develop a variant (marked with an asterisk, e.g., Path Cuts*) that uniformly samples 1\% nodes.
		\item Manifold sampling methods, including the unbiased variant of Specular Manifold Sampling (SMS) \cite{zeltner2020specular} and Manifold Path Guiding (MPG) \cite{fan23mpg}. 
		\item Photon-based (biased) methods, including stochastic progressive photon mapping (SPPM) \cite{Hachisuka09SPPM} and metropolised bidirectional estimator (UPSMCMC) \cite{sikRobustLightTransport2016}. We tune the initial photon lookup radius to lower the bias and only compare noise.
		\item Regular MC methods, including (bidirectional) path tracing \cite{kajiyaRenderingEquation1986, veachBidirectionalEstimatorsLight1995}, path guiding \cite{mullerPracticalPathGuiding2017}, and Metropolis light transport \cite{jakob2012manifold, veachMetropolisLightTransport1997}.
	\end{itemize}
	\note{We also evaluate the temporal stability in equal-time and equal-sample settings in the supplemental video.}

	\paragraph{Comparisons with deterministic search.}
	
	The main drawback of deterministic search is its high computational cost, which leads to extremely low sample rates.
	Consequently, although it produces zero-variance estimations of incident radiance, the rendering result still suffers from aliases and noise. This issue is particularly pronounced for specular-diffuse-specular (SDS) effects, which require path tracing to sample diffuse shading points (e.g., in Fig. \ref{fig_eqtime}, the caustics viewed through the reflection of the gold plane). Besides, there are many fireflies on the floor, which come from the sampling of non-specular paths (with potential connections to the deterministic specular ones).
	Fortunately, our method significantly speeds up rendering by stochastically reducing the search domain using bounding information. As a result, we can utilize substantially higher sample rates within equal time, effectively decreasing the overall noise and aliases present in the final image.
	
	For multiple scattering,  deterministic search becomes too slow due to combinatorial explosions. As a result, we can only uniformly sample a portion (Path Cuts*) for roughly equal-time comparisons. However, uniform sampling does not consider energy, thus introducing significant noise, as shown in Fig. \ref{fig_eqtime2}.
	Our bound-driven sampling reduces the number of tuples that need to perform root-finding. Consequently, rendering with one sample per pixel takes just one to several seconds while maintaining a low variance.
	
	
	\paragraph{Comparisons with manifold sampling.}
	
	As state-of-the-art methods for sampling specular paths, both SMS and MPG rely on point sampling to search for admissible chains. In particular, SMS tends to exhibit noticeable noise, largely due to its uniform sampling of seed chains. While MPG mitigates this issue through importance sampling, it requires a fairly long time to learn accurate distributions. With a limited budget, MPG still produces noisy outputs.
	Moreover, these methods possess an unbounded probability of finding a solution, which, depending on the initialization of distributions, could be quite small and result in extremely high variance.
	
	Although we also introduce stochastic sampling, we can keep the variance controllable thanks to the bound of caustics. Consequently, we guarantee that important solutions can be immediately found with a large enough probability, thus achieving low variance.
	
	Note the speed difference between our triangle-based approach and manifold sampling when handling complex geometry. Specifically, even if we sample multiple solutions per shading point while manifold sampling generates at most one, our sample rates are usually higher.
	The reasons for this phenomenon are two-fold:
	\begin{itemize}
		\item For manifold sampling methods, the use of \textbf{reciprocal probability estimation} contributes a significant amount of variance and overhead, resulting in fireflies. In contrast, our triangle sampling does not require probability estimations.
		\item Manifold walks require tracing a full specular chain in each iteration of Newton's solver, which includes several \textbf{intersection tests} (i.e., querying the BVH). Since our triangle sampling has bounded the domain to a local region, our solver does not require ray tracing except for the visibility check after a solution is found.  
	\end{itemize}
	
	
	\paragraph{Comparison with photon-based methods}
	
	Photons are often distributed non-uniformly across the receiver. Low-energy regions, such as the bottom crop of the Slab scene, may receive an insufficient number of photons. However, accurate density estimations require sufficient photon samples to reconstruct the true distribution reliably. As a consequence, rendering results may exhibit either noise or blurriness, depending on the choice of the kernel radius.
	Our approach operates on functions within finite regions and directly solves for admissible paths, avoiding the issues related to point sampling and density estimations. As a result, we achieve low-variance rendering that preserves the sharp details of caustics.

	\paragraph{Comparison with regular MC methods}
	
\begin{figure}[t]
			\centering
			\begin{minipage}{\linewidth}
				\includegraphics[width=1\linewidth, trim=78 0 0 0, clip]{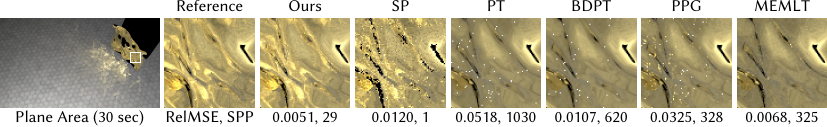}
			\end{minipage}
			\caption{
				\textbf{Equal-time (30 sec) comparisons on handling area light sources}.
				We compare our method with path tracing (PT), bidirectional path tracing (BDPT), Practical Path Guiding (PPG), and Manifold Exploration Metropolis light transport (MEMLT) in a Plane scene lit by an area light. Precomputation takes 11 sec.
			}
			\label{fig_cmp_pt}
		\end{figure}

	Traditional MC methods face significant challenges when dealing with SDS paths (Fig. \ref{fig_cmp_pt}) because of the high-frequency radiance distribution. Even with effective guiding or Metropolis sampling, they still rely on the base sampler to find initial paths for subsequent learning and mutations. As an intrinsic limitation, these issues also persist with more advanced guiding and Metropolis sampling.

	%
	%
	%
	%


		\subsection{Validations}
		
		In addition to the above rendering results, we provide a direct visualization of the correctness and tightness of our bound.
		
		\paragraph{Bounding correctness and tightness} 
		
\begin{figure}[t]
			\centering
			\begin{minipage}{\linewidth}
				\includegraphics[width=\linewidth, trim=0 3 0 3]{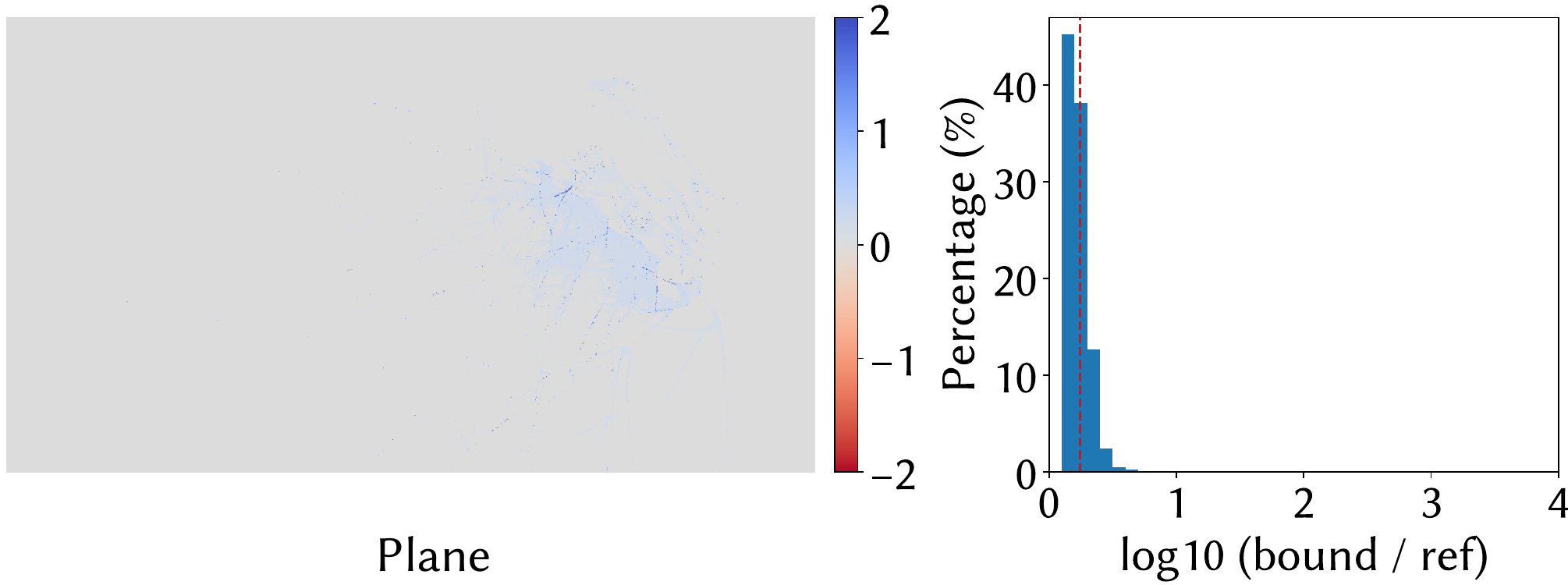}
				\includegraphics[width=\linewidth, trim=0 3 0 3]{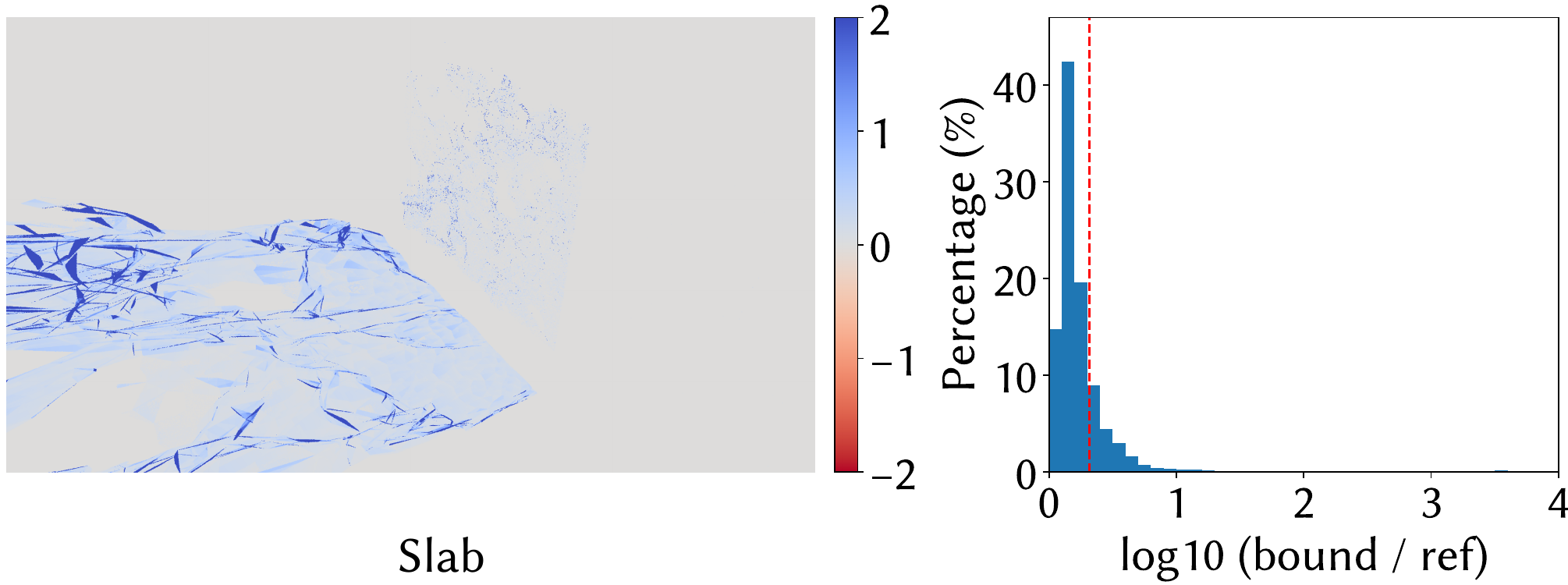}
			\end{minipage}
			\caption{
				\textbf{Visualization of the ratio between the bound and the true irradiance} for each solution in the Plane and Slab scenes. The image illustrates the overall situation, where the ratio is averaged per pixel. The accompanying histogram represents the ratios for each solution. These ratios are displayed on a logarithmic scale with base 10. All ratios are greater than zero, and the red dashed lines indicate the average.
			}
			\label{fig_ratio}
		\end{figure}

		In Fig. \ref{fig_ratio}, we present the ratio \(\tilde E / E\) between the irradiance bound and the true value, shown in a logarithmic scale with base 10. This ratio is expected to be no smaller than 1 as long as the bound is valid, with smaller values representing a tighter bound.
		The absence of red regions in the image indicates that our bound is consistently valid.
		The predominance of light blue areas suggests that the bound is generally tight.
		Note that the bound may become loose for various reasons, including a loose position/irradiance bound and insufficient resolution.
		
		\paragraph{The number of solutions} 
		
		We further validate the number of solutions for each triangle tuple, as illustrated in Fig. \ref{fig_nroots}. Across our tested scenes, nearly all triangle tuples exhibit at most one solution. This indicates that our assumption of $m=1$ is reasonable.

		\begin{figure}[t]
			\centering
			\begin{minipage}{\linewidth}
				\includegraphics[width=1\linewidth, trim=0 7 0 6]{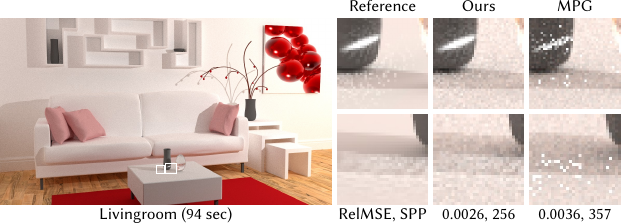}
			\end{minipage}
			\caption{ Rendering scenes with non-planar caustics receivers. \note{Our method is still accurate, though relatively slow.} Precomputation takes 20 sec. 
			}
			\label{fig_multi_rcvr}
		\end{figure}

		\subsection{Ablation studies}
		
		We study the impact of some important components in our pipeline.

		\paragraph{With vs. without remainder variables}
		
\begin{figure}[t]
			\centering
			\begin{minipage}{\linewidth}
				\includegraphics[width=1\linewidth, trim=0 3 0 3]{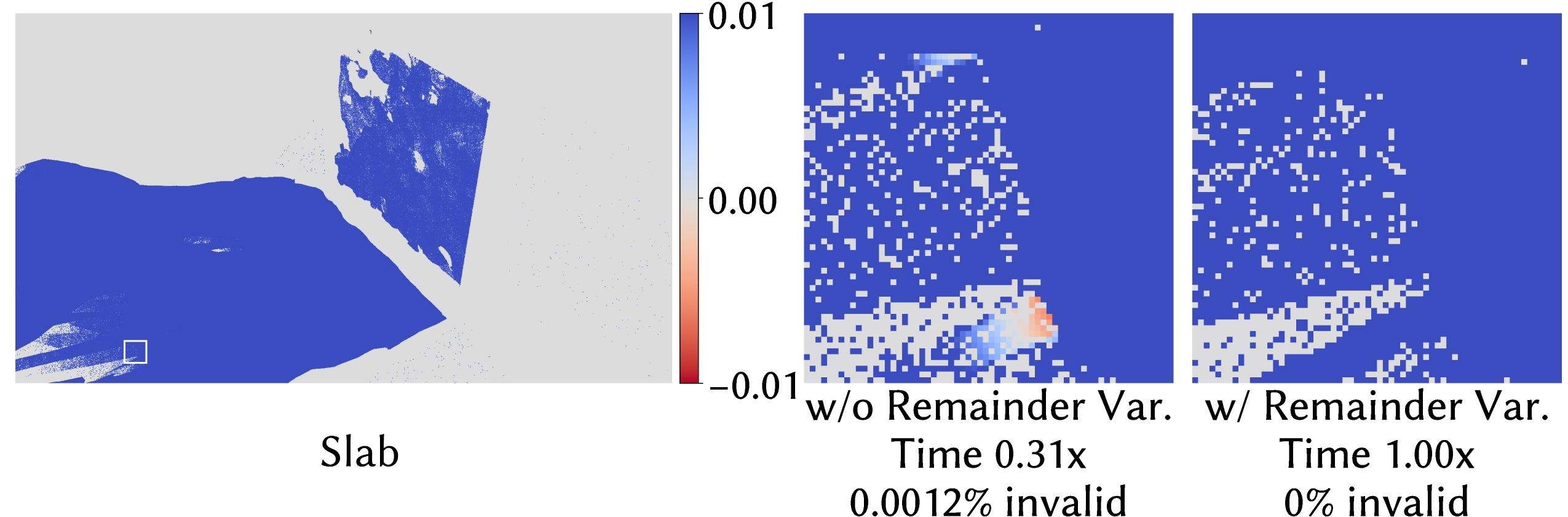}
			\end{minipage}
			\caption{ \textbf{Without remainder variables}, there exists some solutions whose bound $\tilde E$ is slightly lower than the true irradiance $E$ (red). By using remainder variables, all solutions are properly bounded (blue). The ratio $\tilde E / E$ is displayed on a logarithmic scale with base 10.
			}
			\label{fig_ablation_slab}
		\end{figure}

		In Fig. \ref{fig_ablation_slab}, we examine the impact of the remaining variables. The absence of these variables significantly accelerates precomputation; however, this comes at the cost of bounding validity, leading to (red) regions where the irradiance exceeds bounds. By incorporating the remaining variables, we ensure validity, albeit with a slower precomputation.
		
		For our experiments, we enable the remaining variables unless otherwise noted for the sake of strict correctness. Nonetheless, we acknowledge that in certain scenarios, maintaining bounding validity may not be critical. In such cases where slight leaking or increased variance is acceptable, one might consider omitting the remaining variables to enhance performance.

		\paragraph{Position-only vs. our complete method} 
		
		It is possible to compute only the position bound and then either enumerate or uniformly sample all tuples that cover the shading point. We evaluate these variants in Fig. \ref{fig_cmp_sample}. As seen, enumeration yields a significantly smaller number of samples, leading to higher overall noise. Uniform sampling, which does not account for energy, introduces visible variance. Our method, by sampling a small subset based on irradiance bound, accelerates rendering while maintaining low variance, ultimately resulting in high-quality rendering within equal time.
		
\begin{figure}[t]
			\centering
			\begin{minipage}{\linewidth}
				\includegraphics[width=1\linewidth, trim=0 7 0 2]{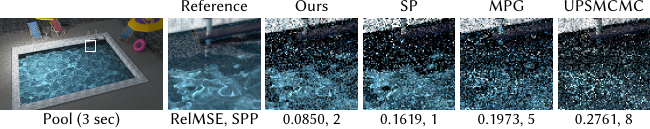}
			\end{minipage}
			\caption{
				The advantage of our method is not significant in cases already well handled by deterministic search. We show an example of a pool scene with shallow water, where the number of tuples related to each shading point is small, and irradiance does not have a significant difference. Precomputation time is included, which takes 1 sec. 
			}
			\label{fig_eqtime3}
		\end{figure}

		However, it is important to note that the effectiveness of irradiance-based sampling is scene-dependent. In scenarios where the number of tuples that cover a grid cell is low (e.g., the Slab scene in Fig. \ref{fig_eqtime2} and the Pool scene in Fig. \ref{fig_eqtime3}), tuple sampling may contribute minimally. In such cases, bypassing the irradiance computation can shorten the precomputation time. Just allocating these budgets to the rendering pass would result in higher overall quality.
		
		\paragraph{Interval arithmetic vs. Bernstein polynomials}
		
		Our pipeline can support various methods for computing the bounds of rational functions. In Fig. \ref{fig_interval}, we compare our approach with interval arithmetic, which has been widely adopted in previous works \cite{walterSingleScatteringRefractive2009, wang2020path}. Despite its general applicability, we observed that interval arithmetic exhibits slow convergence and often generates excessively loose bounds, particularly for irradiance. In contrast, our use of Bernstein polynomials takes advantage of the properties of rational functions, yielding tighter bounds in both scenarios of equal piece count and equal time allocation.

		
		%

		\begin{table*}[t]
			\centering
			\footnotesize	
			\caption{
				\textbf{Rendering statistics}. We show the percentage of \note{precomputation} time used for position bound (including tuple constructions), irradiance bound, and recording bound into grid cells. For rendering, we show the percentage of time used by sampling $\mathrm{S}$ from $\mathrm{U}$  and the average size of $\mathrm{S}$, $\mathrm{B}$, and $\mathrm{U}$, respectively. We only render the specified chain type for fair comparisons. Additionally, we report the number of triangle tuples, the number of pieces (that require computing position bounds and irradiance bounds, respectively) averaged per tuple, and the size of bound storage.
			}
			\begin{tabular}{llllrr|rrr|rrrr|rrrr}
				\toprule
				Figure & \multicolumn{2}{c}{Type} & Scene & \#Tri. & Subdiv. & \multicolumn{3}{c|}{Precomputation Time} & \#Tuples & \multicolumn{2}{c}{Avg. \#Pieces}  & Mem. & Render Time & \multicolumn{3}{c}{Size of sets} \\
				& Chain & Normal & & (K) & Max Level & Pos. & Irr. & Rec. & (K) & Pos. & Irr. & (MB) & Sampling & |S| & |B| & |U| \\
				\midrule
				Fig. \ref{fig_teaser} & R & Interp & Dragon (1) & 354 & 20 & 23.4\% & 58.9\% & 16.5\% & 354.26 & 43.44 & 32.12 & 1948.7 & 3.2\% & 11.8 & 13.0 & 1454.4 \\
				Fig. \ref{fig_teaser} & R & Interp & Dragon (23) & 354 & 20 & 21.5\% & 74.8\% & 2.6\% & 354.26 & 193.19 & 170.49 & 1297.4 & 2.0\% & 1.3 & 2.0 & 1053.7 \\
				Fig. \ref{fig_eqtime} & R     & Interp & Plane     & 131  & 12 & 21.6\% & 55.2\% & 21.8\% & 131.07 & 10.89 & 7.75 & 220.9 & 3.1\% & 3.4 & 4.1 & 204.7 \\
				Fig. \ref{fig_eqtime} & T     & Interp & Sphere     & 82  & 1 & 4.1\% & 92.6\% & 0.7\% & 81.77 & 1.76 & 1.17 & 37.0 & 1.4\% & 1.5 & 1.7 & 51.6 \\
				Fig. \ref{fig_eqtime2} & TT     & Interp & Slab     & 10  & 1 & 20.4\% & 62.4\% & 10.1\% & 126.87 & 1.58 & 0.37 & 28.6 & 1.3\% & 18.7 & 19.5 & 20.5 \\
				Fig. \ref{fig_eqtime2} & TT     & Flat & Diamonds     & 10  & 3 & 4.9\% & 76.3\% & 11.7\% & 192.57 & 3.63 & 0.77 & 29.8 & 6.5\% & 2.5 & 3.3 & 15.1 \\
				Fig. \ref{fig_multi_rcvr} & TT & Flat & Livingroom & 3 & 2 & 2.2\% & 91.8\% & 2.1\% & 48.79 & 3.36 & 0.65 & 4.6 & 0.4\% & 0.7 & 0.8 & 0.8 \\
				Fig. \ref{fig_eqtime3}  & T & Interp & Pool & 20 & 1 & 6.5\% & 84.2\% & 0.9\% & 20.00 & 1.12 & 1.12 & 13.0 & 2.6\% & 3.3 & 3.8 & 7.6 \\
				\bottomrule
			\end{tabular}
			\label{tab:perf}
		\end{table*}
		
		\subsection{Performance analysis}
		
		In Table \ref{tab:perf}, we report the statistics of our rendering experiments. 
		
		\paragraph{Precomputation time}
		In the precomputation pass, the majority of time is spent calculating the irradiance bounds due to the high degrees. The recording process also incurs some overhead because we utilize a simple uniform grid, which becomes inefficient when the position bounds cover numerous cells.
		
		\paragraph{Sampling}
		The size of our sampled set \(\mathrm{S}\) is often smaller than that of \(\mathrm{U}\), which validates the effectiveness of sampling. 
		As we pack tuples into bins whose accumulation of probabilities never exceeds 1, the number of bins $|\mathrm{B}|$ is slightly above \(|\mathrm{S}|\). Sampling a tuple from a bin only requires a simple bisection, while solving for an admissible path is inherently more time-consuming. As a result, the time added by our sampling process is relatively minimal.
		
		\paragraph{The number of triangle tuples and pieces}
		For double scattering, outgoing rays from each \(\mathcal{T}_1\) intersect with only 10 to 20 different \(\mathcal{T}_2\) on average. This indicates that our bounds effectively mitigate the combinatorial explosion associated with triangle-based methods.
		
		The average number of pieces after subdivision remains far below the quartic of the maximum subdivision level, thanks to the stopping criteria for domain subdivisions.

		\paragraph{Impact of mesh tessellations}
		
		We further investigate the relationship between performance and tessellations in Table \ref{tab:subdiv}. 
		Generally, our method effectively keeps rendering time and error at a stable level, albeit with a slight growth as the mesh tessellation increases. Precomputation time and memory usage also grow sublinearly.
		
		We highlight the double refraction case, where a na\"ive combination of triangle tuples would result in a quadratic increase relative to the number of triangles. Thanks to our bound-driven tuple constructions, the growth in the number of tuples is linear to the number of triangles. Note that the bound-driven tuple constructions also enable handling of non-planar caustics receivers as shown in Fig. \ref{fig_multi_rcvr}, with a sublinear (nearly square root) growth of precomputation time with respect to the number of triangles of the receiver.
		Additionally, the average precomputation time and storage requirements per tuple decrease as the position bounds become smaller and the irradiance variation within each tuple is reduced.
		
		In the absence of domain subdivisions (Level = 0), precomputation time, rendering time, and storage costs all increase significantly compared to scenarios that utilize subdivisions. This underscores the necessity of introducing domain subdivisions.

		\paragraph{Impact of hyper-parameters}
		
\begin{figure*}[t]
			\centering
			\begin{minipage}{\linewidth}
				\includegraphics[width=\linewidth, trim=0 85 0 3,clip]{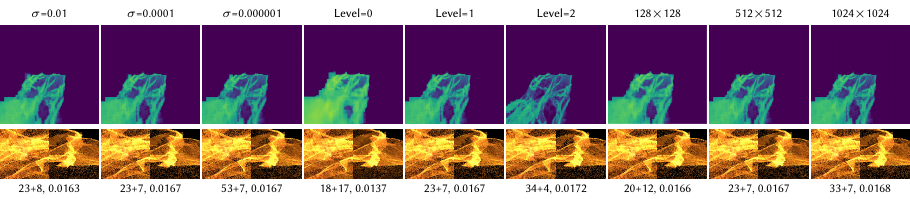}
				\includegraphics[width=\linewidth, trim=0 2 0 30, clip]{par/param_test1.pdf}
			\end{minipage}
			\caption{ 
				\textbf{The influence of precomputation parameters}, including the spatial threshold $\sigma$, the maximal level of subdivisions, and grid resolutions. We visualize the irradiance bound summed over tuples. Precomputation time $p$, rendering time $q$\note{, and RelMSE values $r$} are reported in the form $p+q, r$. 
			}
			\label{fig_par}
		\end{figure*}

		We assess the impact of precomputation parameters in Fig. \ref{fig_par}. Generally, these parameters govern the trade-off between precomputation time and rendering time required to achieve a consistent noise level. For instance, a smaller spatial threshold \(\sigma\), finer subdivision depth, and higher grid resolutions facilitate faster\footnote{The rendering time has a limit proportional to $\gamma$ times the sum of true irradiance.} rendering convergence, albeit at the expense of increased precomputation time. For each scene, our selected parameters strike an appropriate balance in this regard. Yet, we leave the automatic selection strategies for future work.
		Meanwhile, it is important to note that these precomputation parameters have minimal influence on rendering quality, which is predominantly determined by the sampling parameter \(\gamma\), as illustrated in Fig. \ref{fig_par1}.
		
		\begin{table}[t]
			\centering
			\footnotesize
			\caption{\textbf{The impact of mesh tessellation (uniform) on single reflection (top) and double refractions (bottom).} \textbf{Top:} rendering variants of the Plane scene. We observe a sublinear growth of precomputation time, rendering time, and memory with respect to the increase in the number of triangles. We use the fixed maximal subdivision level at 10. \textbf{Bottom:} We show statistics on rendering the double refraction of a ball (with interpolated normal). Here, level refers to the maximal domain subdivision depth, which we decrease as the mesh gets finely tessellated.}
			\begin{tabular}{r|rrrr}
				\toprule
				{\#Triangles/K} & {Precomputation/s} & {Rendering/s} & {Memory/MB} & {RelMSE} \\
				\midrule
				7 & 0.4 & 6.0 & 56.2 & 0.00327 \\
				28 & 0.4 & 6.0 & 89.7 & 0.00359 \\
				114 & 0.7 & 6.8 & 138.5 & 0.00391 \\
				458 & 1.5 & 8.0 & 205.8 & 0.00417 \\
				\bottomrule
			\end{tabular}
			\begin{tabular}{rr|rrrrr}
				\toprule
				{\#Triangles} & {Level} & {\#Tuples/K}  & {Pre./s} & {Render./s} & {Mem./MB} & {RelMSE} \\
				\midrule
				80 & 4 & 15.0 & 4.1 & 3.2 & 22.1 & 0.00007 \\
				320 & 3 & 52.4 & 7.1 & 2.2 & 23.8 & 0.00006 \\
				1280 & 2 & 198.3 & 17.5 & 5.1 & 72.1 & 0.00006 \\
				5120 & 1 & 770.6 & 47.1 & 7.3 & 110.4 & 0.00006 \\
				\midrule
				320 & 0 & 15.0 & 8.8 & 199.9 & 687.1 & 0.00002 \\
				1280 & 0 & 52.4 & 10.9 & 91.0 & 641.1 & 0.00002 \\
				5120 & 0 & 198.3 & 18.2 & 31.6 & 400.9 & 0.00003 \\
				\bottomrule
			\end{tabular}
			\label{tab:subdiv}
		\end{table}

		\section{Limitations}
		
		\paragraph{Convergence of bounds}

		Due to the approximations employed and the necessity of rational functions, \note{our current framework does not yet include a theoretical analysis regarding the guaranteed convergence rate and the tightness of the bound, so our precomputation could be costly.} Practically, in some cases, such as when triangles intersect with each other, the bound is extremely loose, necessitating substantial subdivisions. Establishing a tight and efficient bound for these challenging scenarios deserves future research.

		%
		%
		\paragraph{Generalization with remainder variables}
		
		Our method is currently designed for triangle meshes with a computational cost sublinear to the number of triangles. However, theoretically, this approach is not constrained due to the strong capability of remainder variables in expressing uncertainty and approximation errors. Just like how we handle area light sources, future work could explore extensions to near-specular vertices, non-planar triangles, and triangle aggregations, which could no longer depend on the number of triangles in the scene but rather on the actual geometric complexity.
		
		\paragraph{Long chains}
		
\begin{figure}[t]
			\centering
			\begin{minipage}{\linewidth}
				\includegraphics[width=1\linewidth, trim=0 0 0 0, clip]{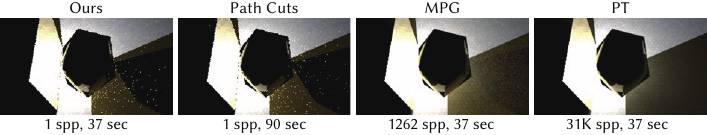}
			\end{minipage}
			\caption{
				\textbf{Triple reflections} between a metallic plane and a sphere. We precompute only the position bounds, which requires 8 sec. Due to the looseness of our bounds, the resulting speedup is not substantial.
			}
			\label{fig_rrr}
		\end{figure}

		Our method is tailored for high-quality rendering of short specular chains (i.e., one or two bounces), where it demonstrates the most substantial improvements over existing techniques. As the length increases, we may encounter additional challenges. Within a triangle tuple, the degree of rational functions becomes higher. Although we can convert them into low-degree ones, it results in looser bounds and an increased computational burden (Fig. \ref{fig_rrr}). More critically, the number of triangle tuples will grow, even with our bound-driven tuple constructions. Therefore, we believe it is impractical to consider all possible triangle tuples during precomputation. Instead, it would be more appropriate to focus on those with high contributions from the beginning.
		
		\note{
			\paragraph{Simplifying assumptions}
			Visibility and the Fresnel term are ignored during precomputation. We also assume a single, small emitter and purely specular scattering. Thus, the precomputation time would scale linearly with the number of emitters. Also, we only consider specular chains connecting to light sources. Future work could relax these assumptions to make the method more practical.
		}
		
		\paragraph{Bound representations}
		
		We parameterize the positional bound using the texture coordinates of receivers and employ a uniform grid for simplicity. This approach is advantageous for planar receivers; however, it faces performance degradation as the complexity of the receivers increases (see Fig. \ref{fig_multi_rcvr}). Future work could store volumetric bounds with vector irradiance \cite{Arvo94}, thereby eliminating dependence on the receiver configuration. Implementing a spatial hierarchy could further reduce computational and memory costs.

		\section{Conclusion}
		
		The challenge of unbounded convergence is crucial to robustly render complex light transport. By bounding both the position and irradiance of caustics, we succeed in controlling the estimator's variance, resulting in efficient and robust rendering. 
		With analytic and functional modeling on both the light transport behaviors and geometric information, we finally reach a bound of caustics using the properties of rational functions. Unlike methods based on point sampling and online learning, our bound is intrinsically reliable and conservative. We finally leverage our bound to achieve a variance reduction of over an order of magnitude in equal time. 
		
		We believe our method represents a step forward in controlling the complex behaviors of stochastic sampling, indicating great potential for efficient and reliable rendering. The established bounds may have further applications beyond triangle sampling, such as manifold sampling and general path guiding. Additionally, we hope our method will inspire future research focused on developing enhanced bounds for caustics and beyond.

		\begin{acks}
			We would like to thank the anonymous reviewers for their valuable suggestions. We also thank Pengpei Hong, Pengcheng Shi, and Huzhiyuan Long for proofreading and valuable discussions.
			
			This work was supported by the National Natural Science Foundation of China (No. 61972194 and No. 62032011) and the Natural Science Foundation of Jiangsu Province (No. BK20211147).
			
			The model in the scene Diamond has been created by the cgtrader user alexmit. The scene Sphere, Slab, Pool, and Plane are modified from the test scenes from \citet{zeltner2020specular}. Other scenes are acquired from \citet{resources16} with modifications. 
		\end{acks}

		\bibliographystyle{ACM-Reference-Format}
		\bibliography{mainref}

		%
		%

		%
		%
		%
		%
		%
		%
		%
		%
		%
		%
		%
		%
		
		\appendix

\end{document}